\documentclass[twocolumn, aps, prx, nofootinbib]{revtex4-1}
\usepackage{graphicx}
\usepackage{amsfonts}
\usepackage{amssymb}
\usepackage{amsthm}
\usepackage{array}
\usepackage{amsmath}
\usepackage{verbatim} 
\usepackage{hyperref}
\usepackage{color}
\usepackage{bbold}
\usepackage{epstopdf}
\usepackage{mathtools}
\usepackage[normalem]{ulem}
\usepackage{accents}
\usepackage{scalerel}

\newtheorem{definition}{Definition}
\newtheorem{theorem}{Theorem}
\newtheorem{corollary}{Corollary}[theorem]

\newcommand{\norm}[2]{\left\Vert {#1} \right\Vert_{#2}}
\newcommand{\ket}[1]{\left\vert#1\right\rangle}
\newcommand{\bra}[1]{\left\langle#1\right\vert}

\def\bra#1{\langle #1|}
\def\ket#1{ | #1 \rangle}

\def\JJ#1#2{{\cal J}_{#1}^{#2}}
\def\Tr{\mbox{\rm Tr}}

\begin{document}
\title{Fluctuation Theorems for a Quantum Channel}
\author{Hyukjoon Kwon}
\email{h.kwon@imperial.ac.uk}
\author{M. S. Kim}
\email{m.kim@imperial.ac.uk}
\affiliation{QOLS, Blackett Laboratory, Imperial College London, London SW7 2AZ, United Kingdom}
\begin{abstract}
We establish the general framework of quantum fluctuation theorems by finding the symmetry between the forward and backward transitions of any given quantum channel. The Petz recovery map is adopted as the reverse quantum channel, and the notion of entropy production in thermodynamics is extended to the quantum regime. Our result shows that the fluctuation theorems, which are normally considered for thermodynamic processes, can be a powerful tool to study the detailed statistics of quantum systems as well as the effect of coherence transfer in an arbitrary non-equilibrium quantum process. We introduce a complex-valued entropy production to fully understand the relation between the forward and backward processes through the quantum channel. We find the physical meaning of the imaginary part of entropy production to witness the broken symmetry of the quantum channel. We also show that the imaginary part plays a crucial role in deriving the second law from the quantum fluctuation theorem. The dissipation and fluctuation of various quantum resources including quantum free energy, asymmetry and entanglement can be coherently understood in our unified framework. Our fluctuation theorem can be applied to a wide range of physical systems and dynamics to query the reversibility of a quantum state for the given quantum processing channel involving coherence and entanglement.
\end{abstract}
\maketitle

\section{Introduction}
Since Shannon's adaptation of `entropy' as the measure of information \cite{1}, we have witnessed surprising usefulness of the mathematical descriptions of thermodynamics in developing information theory \cite{2}. The interplay between information theory and thermodynamics has gone further than the development of common mathematical tools as the information \cite{3,4} and its feedback control \cite{Sagawa10} were, indeed, found to be physical and  a source of work. 

Any physical processes can involve inevitable loss, and this gives rise to the irreversibility of macroscopic states, which is known as the second law of thermodynamics. The principle of macroscopic irreversibility is not only valid in thermodynamics, but also in information theory. A resembling theorem called the data processing inequality, which says that information content never increases through a noisy channel, presides in information theory. These laws and theorems are, however, based only on the average behavior of a large system in equilibrium, while the full picture of statistical properties of a physical process can be found through the dynamics of probabilities of microstates. Macroscopic physical states consist of the ensemble of possible microstates, and physical processes can be understood as a collection of transitions between microscopic states.  As the physical system gets smaller and more complex, there has been a demand to describe such transitions in non-equilibrium. For this purpose, fluctuation theorems (FTs) \cite{Gallavotti95,Crooks99,Jarzynski97,Campisi11} have emerged. In particular, the Crooks FT \cite{Crooks99} can be summarized by the following equality between the probabilities $P_\rightarrow$ and $P_\leftarrow$ of forward and backward transitions 
\begin{equation}
\label{CFTEq}
\frac{P_\rightarrow (\sigma)}{P_\leftarrow (-\sigma)} = e^{\sigma},
\end{equation}
where the parameter involved is the entropy production $\sigma$. This symmetry shows that the probability of the reverse process happening depends on the entropy production during the forward process, which can be understood as a general result from the microreversibility of Hamiltonian dynamics \cite{Maes03}. The Jarzynski equality \cite{Jarzynski97} and the second law of thermodynamics can be derived from the Crooks FT. The classical FTs are well-established with experimental verifications in microscopic systems \cite{Liphardt02,Colin05, Bick'l'e06, Saira12}, in which the transition probability of a microscopic state gives critical effects on the system.

As the theory of thermodynamics extends its realm to the quantum regime, one of the important questions is whether and how the FTs can be generalized to quantum systems. In other words, the question is `Can we establish a single-parameter valued symmetry between forward and backward probabilities for an open quantum process?' In this paper, we answer this question by establishing such a symmetry, which can be applied not only to quantum thermodynamic channels but also to any noisy quantum information processing channels. 

Despite the efforts to extend the FTs to the quantum regime \cite{Esposito09, Campisi11, Hanggi15}, a quantum FT (QFT), which can fully incorporate quantum features in the system and the channel, is still to emerge. For instance, in the presence of quantum coherences, thermodynamic free energy \cite{Lostaglio15, Cwiklinski15, Brandao15} can be larger than its classical counterpart which is concerned with classical energetic values. The role of quantum coherences and correlations is acknowledged as a resource that can be utilized for work extraction \cite{Aberg14, Perarnau-Llobet15, Korzekwa16} or time referencing tasks \cite{Gour17, Kwon18}. These nonclassical features stemming from quantum coherences not only affect thermodynamic quantities on average but also make the outcome probability distributions differ from classical theory. Quantum coherences are present in non-equilibrium quantum processes \cite{Talkner07, Kafri12, Albash13, Hekking13, Rastegin13, Dorner13, Allahverdyan14, Roncaglia14, Rastegin14, Solinas15, Jarzynski15, Binder15, Goold16, Alhambra16, Talkner16, Alonso16, Cuetara16, Gardas16, Deffner16, Park17, Perarnau-Llobet17, Alhambra18, Lostaglio18, Funo18, Aberg18, Manzano18, Holmes18,Morris18,Bartolotta18}. 

In order to establish a QFT, we have to define (1) the reverse process and (2) the quantum version of entropy production. These are not straightforward tasks, and efforts so far have been mainly to use the classical definition of work for quantum systems \cite{Batalhao14, An15} or to adopt quantum measurements \cite{Roncaglia14}, namely, positive-operator valued measurements (POVMs), in order to consider work based on the two-point measurement (TPM) \cite{Talkner07}. The entropy in classical theory does not reflect quantum coherences; thus the quantum parameter equivalent to the entropy production will be based on the quantum measure of fluctuations. Some progress to take into account coherences in QFT has been made by adopting the techniques of quantum information theory \cite{Kafri12, Rastegin13, Rastegin14, Alhambra18, Aberg18, Manzano18} and quantum field theory \cite{Bartolotta18}, as well as the quantum jump approach \cite{Hekking13} and the master equation approach \cite{Cuetara16}. However, some of them \cite{Hekking13, Rastegin13, Cuetara16, Aberg18,Alhambra18} are limited to specific quantum channels, and their measurement-based approaches \cite{Kafri12, Hekking13, Cuetara16, Manzano18} suffer from the loss of coherences after measurements.

In this paper, we establish the fluctuation relation for a linear quantum channel that reproduces the FTs in the classical thermodynamic limit. This can be done by adopting the reverse quantum process, known as the Petz recovery map \cite{Petz86} and generalizing the concept of entropy production to take into account coherences in a quantum system. We investigate the effect of coherences in a quantum state that makes the fluctuation relation, given by the ratio between the forward and backward transition probabilities, different from the conventional FTs. When the quantum channel induces coherence transfers, the transition between diagonal and off-diagonal elements in the density matrix of a quantum state can be understood via complex-valued quantum entropy production. The emergence of imaginary entropy production is related to the symmetry breaking property of the quantum channel, and we study concrete examples of a two-level atom interacting with coherent and incoherent bath states. More importantly, the imaginary part of entropy production plays an essential role in deriving the generalized second law for a quantum channel from our QFT. Our result verifies that not only the loss of thermodynamic free energy but also the loss of coherences or entanglement can be qualified as a dissipated resource responsible for the irreversibility of a quantum process.

The rest of the paper is organized as follows. In Section \ref{Sec-Prelim}, we reformulate the FT for a classical channel and discuss how to identify the reverse process for a given quantum channel via the Petz recovery map. In Section~\ref{Sec-FTCohState}, we demonstrate how coherences in a quantum state affect the fluctuation relation deviating from conventional FTs and construct the QFT of entropy production based on it. In Section~\ref{Sec-CohChannel}, we introduce the complex-valued quantum entropy production in order to fully describe the transition between off-diagonal elements through the quantum channel and discuss how the QFT should be modified accordingly. In Section~\ref{Sec-2ndLaw}, we show that  the generalized second law of thermodynamics can be derived from the QFT, which can be applied to analyze the loss in quantum thermodynamics, as well as in the resource theories of asymmetry and entanglement. The paper is concluded with final remarks in Section~\ref{Sec-Conclusion}.

\section{Preliminaries}
\label{Sec-Prelim}
\subsection{Entropy production fluctuation relation for a classical channel}
We reformulate the Crooks FT \cite{Crooks99}  in Eq.~(\ref{CFTEq})  which will help us to establish a QFT in a close analogy to this formulation. For this purpose, we focus on the transition between two physical states from $A$ to $B$ by a general physical process ${\cal N}$, which can be a thermodynamic process or information encoding through a noisy channel. The physical states $A = \{ p(x) \}$ and $B = \{ p'(y') \}$ are assumed to be macroscopic states  described by probability distributions $p(x)$ and $p'(y')$ of their microscopic entities $x$ and $y'$. We may consider the reverse process ${\cal R}$ corresponding to ${\cal N}$, which can be achieved by another physical process, for example, time-reversal, information decoding, or a recovery channel.

The transition probability from a microscopic state $x$ to another microscopic state $y'$ for the forward process $x \xrightarrow{\cal N} y'$ is denoted by $T( x \rightarrow y')$. Similarly, the transition probability for the backward (or reverse) process $x \xleftarrow{\cal R} y'$ is denoted by $\tilde T ( x \leftarrow y')$. The ratio between the forward and backward transition probabilities shows the tendency of the microscopic state transitions. Based on this observation, we denote this ratio after taking the logarithm as ``information exchange":
\begin{equation}
\label{EntFluxDef}
\delta q_{x \rightarrow y'} := - \log \left[ \frac{T(x \rightarrow y')}{\tilde T( x \leftarrow y')}\right].
\end{equation}
Note that the information exchange does not depend on the distribution of microstates in the macrostate $A$ or $B$, but on the forward and backward physical processes. Here, we find the concrete physical meaning of information exchange in thermodynamics when a system is in contact with a thermal bath in equilibrium. By assuming the microscopic reversibility with the energy conservation law, the transition probability between two phase space points $x$ and $y'$ are given by $ T(x \rightarrow y') =\bar T ( \bar x \leftarrow \bar y') e^{-\beta \Delta E}$ \cite{Crooks99, Groot62, Chandler87}, where $\bar T$ represents the time-reversed trajectory of the time-reversed states $\bar x$ and $\bar y'$, $\Delta E$ is the energy difference between the two phase space points, and $\beta$ is the inverse temperature: $\beta = 1/ (k_B T)$ with $k_B$ the Boltzmann constant. When the system and bath are isolated from the external environment, the energy difference $\Delta E$ of the system comes from the heat $Q$: $\Delta E = Q$. Thus, the information exchange $\delta q_{x \rightarrow y'}$ corresponds to $\beta Q$. However, correlations between the system and memory, which transfers information to a different time \cite{Sagawa09}, can be an additional parameter involved in $\delta q$ which leads to the modification of the FTs \cite{Sagawa10, Sagawa12} for nonequilibrium thermodynamics of measurements and feedback controls. In this paper, we show that the information exchange can have various physical forms---energy, coherence, or entanglement---depending on the physical process and how its reverse process is constructed.

Next, we discuss how the distribution of microstates can be changed as a result of the physical process. For this, we define the difference between the information contents of the macrostates, using the single-shot entropy difference 
$$
\delta s_{x \rightarrow y'} := -\log p'(y') + \log p(x)
$$
provided that we observe the statistics of the microstates $x$ and $y'$ at the initial and final points. In contrast to the information exchange, this entropy difference depends only on the initial and final probability distributions $p(x)$ and $p'(y')$.
\begin{figure}[t]
\includegraphics[width=0.8\linewidth]{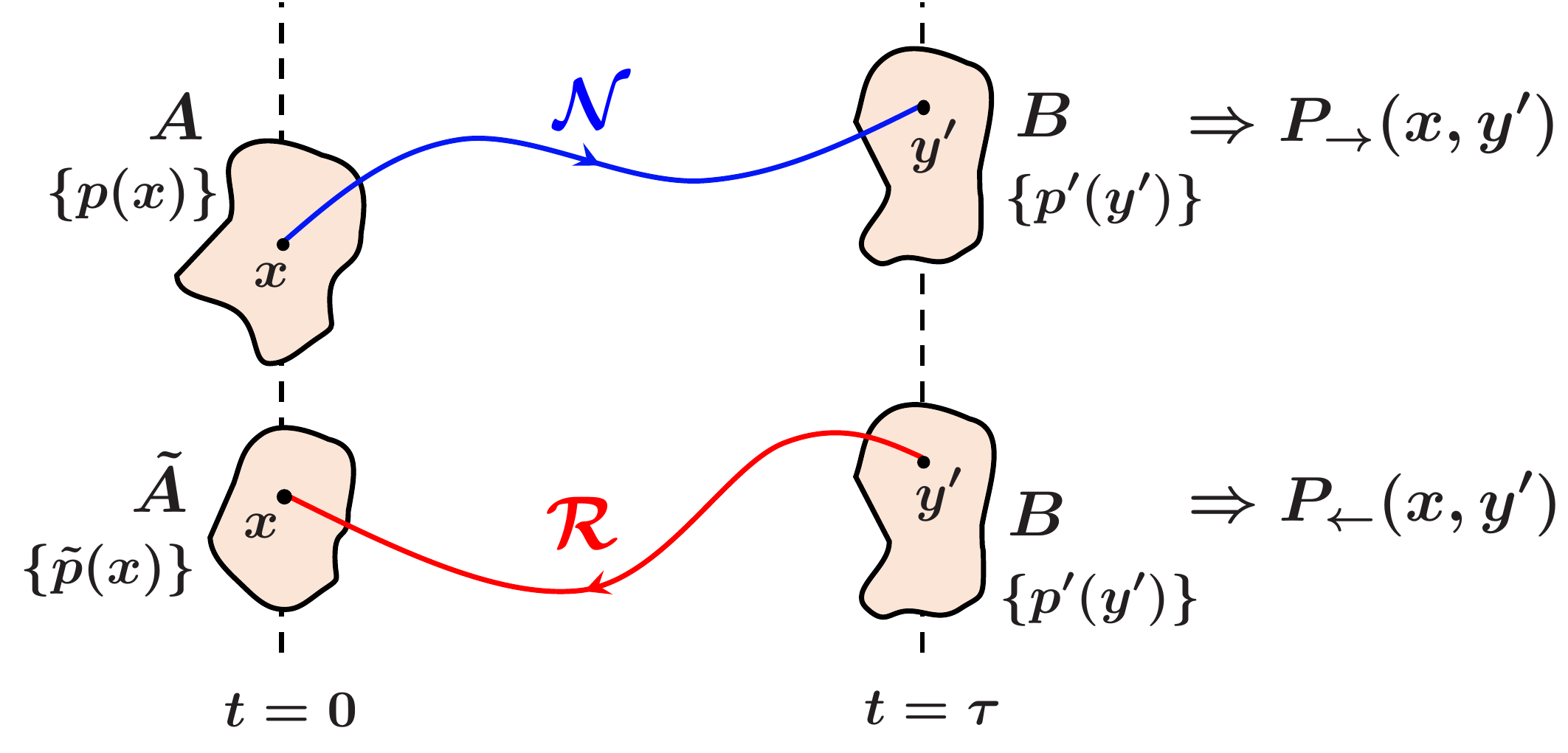}
\caption{Schematic for the standard TPM scheme for the forward $A \xrightarrow{\cal N} B$ and backward $\tilde A \xleftarrow{\cal R} B$ processes. The distributions $P_\rightarrow(x, y')$ and $P_\leftarrow(x,y')$ are defined in the TPM between two different times $0$ and $\tau$.}
\label{TPMFig}
\end{figure}

Finally, we describe how two different entropic quantities---the information exchange and the entropy difference can be connected. Analogous to the thermodynamic entropy production, we define the single-shot entropy production for the transition $x \rightarrow y'$ as
\begin{equation}
\label{CEntProdDef}
\sigma_{x \rightarrow y'} := \delta s_{x \rightarrow y'} - \delta q_{x \rightarrow y'}.
\end{equation}
When the system is in equilibrium, the entropy flow $\delta q_{x \rightarrow y'}$ will lead to the entropy difference $\delta s_{x \rightarrow y'}$, i.e. the entropy production becomes zero: $\sigma_{x \rightarrow y'} = 0$. On the other hand, in the general case of nonequilibrium processes, the entropy production will not vanish. To describe this case, we introduce the TPM approach which is widely studied for FTs in thermodynamics (see Fig.~\ref{TPMFig}). The TPM distribution $P_\rightarrow (x,y')$ describes every possible event to observe $(x,y')$ at the initial and final measurements, where each marginal distribution reduces to the initial or final statistics $p(x)$ or $p'(y')$. Assuming that the physical process depends only on its current state, the TPM joint measurement probability can be written as
$$
P_\rightarrow (x,y') = p(x) T(x \rightarrow y').
$$
Similarly, the backward TPM distribution $P_\leftarrow (x,y') = p'(y') \tilde T( x \leftarrow y')$ can be defined for the backward process. By using the TPM probability distribution, the probability to get $\sigma$ amount of entropy production during the forward process can be written as
$$
P_\rightarrow(\sigma) = \sum_{x,y'} P_\rightarrow (x,y') \delta (\sigma - \sigma_{x\rightarrow y'}),
$$
then we obtain the FT in Eq.~(\ref{CFTEq}). It can be derived from Eq.~(\ref{CFTEq}) that the average entropy production $\langle \sigma \rangle$ is always positive, i.e. ``information loss" during the physical process ${\cal N}$ results in the extra increase of entropy of the system. 

\subsection{Quantum operation time reversal and the Petz recovery map}
In order to generalize the concept of the entropy production FT to a quantum channel, the first task is to identify the reverse map for the given quantum channel, which generalizes the time-reversed trajectory in thermodynamics. We require that both the forward $({\cal N})$ and backward $({\cal R})$ quantum processes should be completely positive trace-preserving (CPTP) maps, which can be expressed as ${\cal N} (\rho) = \sum_m K_m \rho K_m^\dagger$ and ${\cal R}(\rho) = \sum_m \tilde K_m \rho \tilde K_m^\dagger$ satisfying $\sum_m K_m^\dagger K_m = \mathbb 1$ and $\sum_m \tilde K_m^\dagger \tilde K_m = \mathbb 1$ in their Kraus representations.
A form of ${\cal R}$ has been introduced by Crooks \cite{Crooks08} in the notion of `quantum operation time reversal' inspired by the time reversal of the Markov chain \cite{Norris97}. When the channel has a fixed point $\pi$ satisfying ${\cal N}(\pi) = \pi$, we can define the Kraus operator for the reverse dynamics as $\tilde{K}_m = \pi^{1/2} K_m^\dagger \pi^{-1/2}$, which leads to $\Tr [ K_{m'} K_m \pi K_m^\dagger K_{m'}^\dagger] = \Tr [ \tilde{K}_m \tilde{K}_{m'} \pi \tilde{K}_{m'}^\dagger \tilde{K}_m^\dagger  ]$, or equivalently $P_\rightarrow (m, m') = P_\leftarrow (m, m')$, preserving the dynamic history of the process for the state $\pi$. Consequently, the reverse map can be defined as ${\cal R}_\pi(\rho) = \sum_m \tilde{K}_m \rho \tilde{K}_m^\dagger$, where the fixed point $\pi$ remains unchanged by ${\cal R}_\pi$, i.e., ${\cal R}_\pi (\pi) = \pi$.

Crooks's original approach \cite{Crooks08} requires a fixed equilibrium state which is invariant under the quantum channel. This condition is relaxed to construct a general form of the reversed quantum operation, the so-called Petz recovery map \cite{Petz86}. Provided the structure of the quantum channel, i.e., $\gamma \xrightarrow{\cal N} {\cal N}(\gamma)$, is known, the Petz recovery map can be constructed as follows:
\begin{definition} [Petz recovery map] For a given reference state $\gamma$ and CPTP map ${\cal N}$, the Petz recovery map ${\cal R}_\gamma$ is defined as
$$
{\cal R}_\gamma (\rho ) := \left( {\cal J}_\gamma^{\frac{1}{2}} \circ {\cal N}^\dagger \circ {\cal J}_{{\cal N}(\gamma)}^{-\frac{1}{2}} \right) (\rho),
$$
where $  {\cal N}^\dagger( \cdot) = \sum_m K_m^\dagger(\cdot) K_m$ is the adjoint map, and $\JJ{A}{\alpha} (\cdot) := A^\alpha (\cdot) A^{\alpha \dagger}$ is defined as a rescaling map. 
The Petz recovery map is a CPTP map and fully recovers the reference state, i.e. ${\cal R}_\gamma ({\cal N}(\gamma)) = \gamma$.
\end{definition}
In the Kraus representation, the reverse quantum channel ${\cal R}_\gamma$ is given by the set of Kraus operators $\{ \tilde K_m \}$ with $\tilde K_m = \gamma^{1/2} K_m^\dagger {\cal N}(\gamma)^{-1/2}$. Compared with Crooks's quantum operation time reversal \cite{Crooks08}, it is always possible to construct the Petz recovery map ${\cal R}_\gamma$ for any given reference state $\gamma$, while ${\cal R}_\gamma$ reduces to the quantum operation time reversal when taking the fixed reference state $\pi$ satisfying ${\cal N}(\pi) = \pi$. The recovery map ${\cal R}_\gamma$ is specific to the choice of the reference state and it is not possible to choose a map which can recover any initial state $\rho$ unless the channel is represented by a unitary operation for the system. We also note that there exists a duality between ${\cal N}$ and ${\cal R}_\gamma$ that ${\cal N}$ becomes the recovery map of ${\cal R}_\gamma$ by choosing the reference state ${\cal N}(\gamma)$. The reference state can be chosen depending on the fluctuation of which physical properties we are interested in. The recovery maps, following the reference states, can vary even for the same forward quantum channel. As an example, we investigate in Section \ref{Sec-2ndLaw} how the different choices of reference states lead to the different fluctuation theorems of free energy, coherence, or entanglement.

While there have been some discussions of using the Petz recovery map to the QFT \cite{Manzano15, Aberg18, Manzano18, Alhambra18}, they mainly focus on thermodynamic channels \cite{Aberg18, Alhambra18}, or their measurement-based approaches \cite{Manzano15, Manzano18} cause the inevitable loss of coherences after measurements. Here we formulate a very general QFT based on the Petz recovery map. For this, we choose a reference state and its recovery map then write the backward transition probability through this map. In our formulation, both quantum and classical information quantities as well as thermodynamic quantities can be coherently combined into a unified framework. It is worth noting that the dynamics of the system may not be linear, for example when the system and bath are initially correlated  \cite{Romero04}, a case we do not consider in this work. Throughout the paper, we assume that the reference state is full-rank, and we denote the reference state by $\gamma = \sum_i r_i \ket{i}\bra{i}$ and the evolved state by ${\cal N}(\gamma) = \sum_{k'} r'_{k'} \ket{k'} \bra{k'}$, using their eigenvalue decompositions.

\section{Fluctuation theorems for a quantum state with coherence}
\label{Sec-FTCohState}
\subsection{Pure state fluctuation relation}
Let us start with how the fluctuation relation for the transition probability is modified when quantum states contain coherences. Consider the transition probability between two pure quantum states $\ket{\psi} \xrightarrow{\cal N} \ket{\phi'}$, which can be compared to the transition between the microscopic entities $x$ and $y'$ in Eq.~(\ref{EntFluxDef}). Throughout the paper, a primed parameter $(\cdot)'$ denotes the final state after passing the channel. The forward transition probability is defined as $T(\ket{\psi} \rightarrow \ket{\phi'}) := \langle \phi'  |  {\cal N} ( | \psi \rangle \langle \psi | ) | \phi' \rangle$. However, it is important to note that the reference states are not, in general, equally distributed in their eigenstates. Consequently, the maximally mixed state $\mathbb{1}/d$ in a $d$-dimensional Hilbert space is not a passive state as ${\cal N}(\mathbb{1}) \neq \mathbb{1}$ and ${\cal R}_\gamma (\mathbb{1}) \neq \mathbb{1}$. This raise the difficulty of a fair comparison between the forward and backward transitions in the same scale. In order to handle this, the coefficients of a density matrix can be weighted differently based on the distribution of the reference states, which we call the reference-rescaling \cite{Egloff15, Holmes18, Mingo18}.  In particular, we can choose the rescaling operations (denoted as $\sim$) for the reverse process
to satisfy the relation $\frac{\mathbb{1}}{d} \xrightarrow{\sim} {\cal N}(\gamma) \xrightarrow{{\cal R}_\gamma} \gamma \xrightarrow{\sim} \frac{\mathbb{1}}{d}$ so that the maximally mixed states is now mapped to itself in the rescaled statistics. By imposing this condition, the reverse process is applied to the rescaled states  $|\tilde{\psi} \rangle $ and $| \tilde \phi' \rangle$, where the transition probability is given by $\tilde{T} ( | \tilde{\psi} \rangle \leftarrow | \tilde{\phi'} \rangle ) := \bra{\tilde{\psi}} {\cal R}_\gamma  ( | \tilde{\phi'} \rangle  \bra{\tilde {\phi'}}) | \tilde{\psi} \rangle$. We then obtain the following fluctuation relation for the transition between the two pure quantum states:
\begin{theorem}[Detailed balance condition for pure states] The transition probabilities from $\ket{\psi}$ to $\ket{\phi'}$ under a quantum channel ${\cal N}$ and its backward process ${\cal R}_\gamma$ obey the following relation:
\label{PureFT}
\begin{equation}
\label{PureFTEq}
\frac{T( \ket{\psi} \rightarrow \ket{\phi'})}{\tilde{T} ( | \tilde{\psi} \rangle   \leftarrow | \tilde\phi' \rangle) } = \bra{\psi} \gamma^{-1} \ket{\psi} {\bra{\phi'} {\cal N}(\gamma) \ket{\phi'}},
\end{equation}
where $| \tilde{\psi} \rangle = \bra\psi \gamma^{-1} \ket{\psi}^{-1/2} \gamma^{-1/2} \ket{\psi}$ and $| \tilde{\phi}' \rangle = \bra{\phi'} {\cal N}(\gamma) \ket{\phi'}^{-1/2} {\cal N}( \gamma)^{1/2} \ket{\phi'}$ are the reference-rescaled states.
\end{theorem}
\noindent Detailed proofs of all the Theorems can be found in the Appendix. We note that the detailed balance condition given by Eq.~(\ref{PureFTEq}) does not depend on how the quantum channel ${\cal N}$ is applied (for instance, suddenly or adiabatically), which is also the case in classical thermodynamics.  We point out that the eigenstates of the reference states are not affected by the rescaling i.e. $| \tilde i \rangle = \ket{i} $ and $| \tilde k' \rangle = \ket{k'}$. By taking the fixed equilibrium state $\gamma = \sum_i e^{-\beta E_i}/Z \ket{i} \bra{i}$ with the partition function $Z=\sum_i e^{-\beta E_i}$, we obtain $ T( \ket{i} \rightarrow | k' \rangle ) = \tilde T ( \ket{i} \leftarrow | k' \rangle) e^{-\beta \Delta E}$ with $\Delta E = E_{k'} - E_i$, which shows that Eq.~(\ref{PureFTEq}) reduces to the classical detailed balance condition. On the other hand, when the final state is given by the maximally coherent state $\ket{\phi} \propto \sum_{k'} \ket{k'}$, its rescaled state becomes $\ket{\tilde\phi} \propto \sum_{k'} e^{-\beta E_{k'}} \ket{k'}$, the so-called coherent Gibbs state. Conversely, the initial coherent Gibbs state  $\ket{\psi} \propto \sum_{i} e^{-\beta E_i} \ket{i}$ is rescaled into the maximally coherent state $\ket{\tilde{\psi}} \propto \sum_i \ket{i} $ for the reverse quantum process ${\cal R}_\gamma$. Theorem~\ref{PureFT} also shows that recently studied Gibbs-rescaling approaches toward the quantum fluctuation relation \cite{Mingo18, Holmes18} can be applied to a wider range of quantum channels beyond thermodynamic processes.

\begin{figure*}[t]
\includegraphics[width=.95\linewidth]{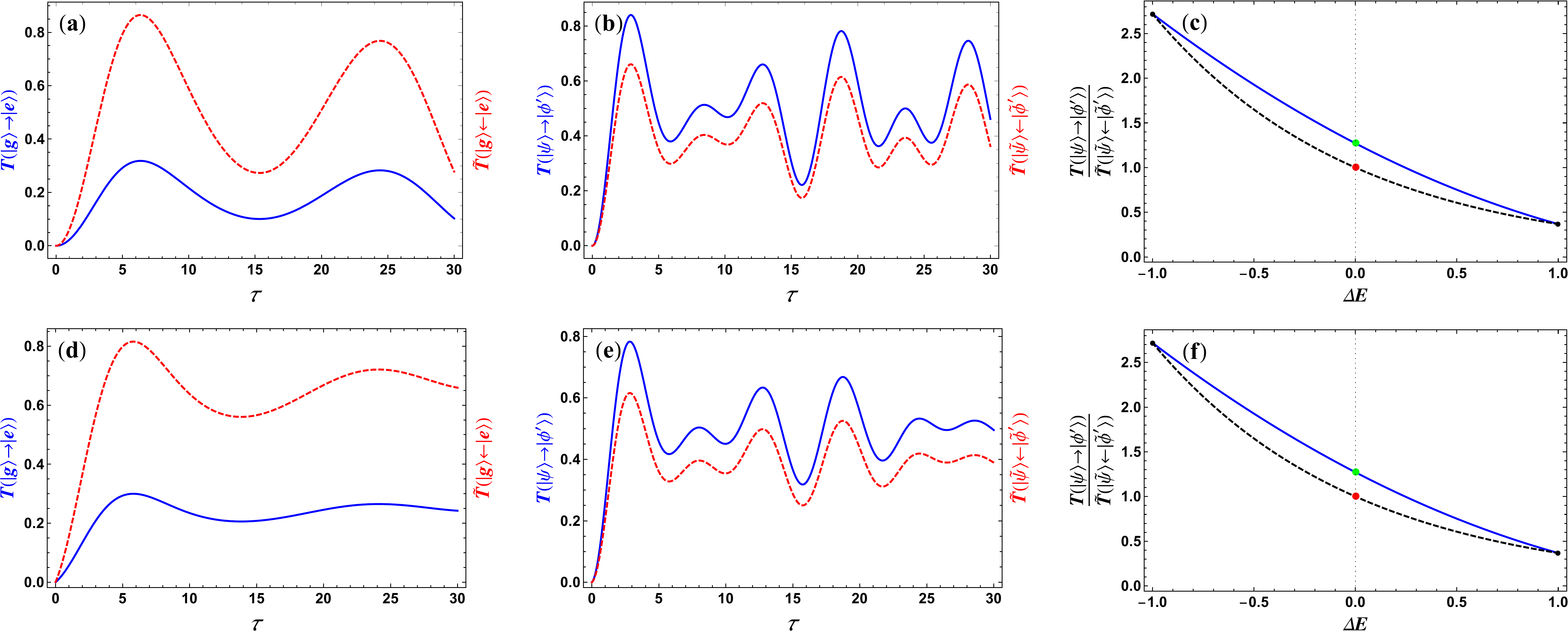}
\caption{Detailed balance of pure states for the JC Hamiltonian described in Eq.~(\ref{JC-h}) with the parameters $\beta = 1$, $\omega_0 =1$ and $g=0.1$. The upper figures (a), (b), and (c) represent the results without noise, and the lower figures  (d), (e), and (f) represent the results with noise  ${\cal L}_{\rm noise}$ given by Eq.~(\ref{JC-diss}) with $\Gamma = 0.1$. (a) and (d) refer to the forward (blue solid lines) and backward (red dashed lines) transition probabilities of $\ket{g} \rightarrow \ket{e}$ versus the evolution time $\tau$. (b) and (e) refer to the state transition $\ket{g} + \ket{e} \rightarrow \ket{g} - \ket{e}$. (c) and (f) show the ratio ${T ( \ket{\psi} \rightarrow \ket{\phi'} ) }/ \tilde {T}( | \tilde{\psi} \rangle \leftarrow | \tilde{\phi'} \rangle)$ between the forward and backward transition probabilities from $\ket{\psi} = \sqrt{r} \ket{g} + \sqrt{1-r} \ket{e}$ to $\ket{\phi'} = \sqrt{1-r} \ket{g} - \sqrt{r} \ket{e}$ versus energy difference $\Delta E = E_{\phi'} - E_\psi$ (blue solid lines). Dashed lines refer to the classical thermodynamic detailed balance $e^{-\beta \Delta E}$. The green and red dots are obtained for the transition with $\Delta E = 0$ for the coherent and incoherent initial fields, respectively. The additional factor $\Upsilon \approx 1.27$ can be observed for the case with coherence. This ratio depends only on the values of $\beta$ and $\omega_0$ but not on $g$, $\Gamma$, or $\tau$.}
\label{PureQFTFig}
\end{figure*}

In order to illustrate the effect of coherence in the detailed balance, let us consider a single two-level atom as our system of interest. This interacts with a simple bath of a single-mode field. The system-bath interaction follows the  Jaynes-Cummings (JC) model \cite{Knight} whose Hamiltonian is given by
\begin{equation}
H_{\rm JC} = H_a +H_f \  + g (\sigma_+ a + \sigma_- a^\dagger),
\label{JC-h}
\end{equation}
where the atomic Hamiltonian $H_a=\hbar \omega_a \sigma_z/2$ with the Pauli $\sigma_z$ operator, the field Hamiltonian $H_f=\hbar \omega_f a^\dagger a$ with bosonic operators $a$ and $a^\dagger$  and the last term represents the interaction Hamiltonian with $\sigma_\pm$ the raising and lowering operators for the atom, and $g$ the atom-field coupling strength. For the resonant interaction, the atomic transition frequency $\omega_a$ and the field frequency $\omega_f$  are the same: $\omega_a = \omega_f = \omega_0$. We assume that the field state is initially in thermal equilibrium at temperature $T$; its density matrix is given by the field density operator $\gamma_f=\exp(-\beta H_f) /Z_f$.  After their interaction for some time $\tau$, the atomic state is found by tracing out the field mode:
\begin{equation}
\label{JCChannel}
{\cal N}_{0 \rightarrow \tau}(\rho_0) = \Tr_{f} [U_{0 \rightarrow \tau} (\rho_0 \otimes \gamma_f) U^\dagger_{0 \rightarrow \tau}],
\end{equation}
where the evolution operator $U_{0\rightarrow \tau} = e^{-i H_{\rm JC} \tau}$. 
As a reference state of the system of interest, let us consider the atomic state in thermal equilibrium, $\gamma_a = \exp(- \beta H_a) / Z_a$, which is unchanged during the time evolution, i.e. ${\cal N}_{0 \rightarrow \tau} (\gamma_a) = \gamma_a$ for any time period $\tau$. This resembles the thermodynamic processes studied in Ref.~\cite{Aberg18} as the unitary evolution obeys the energy conservation relation $[U_{0 \rightarrow \tau}, H_a + H_f] = 0$ .
The corresponding Petz recovery map is written in the form of the time-reversed evolution \cite{Aberg18} of the atomic state
\begin{equation}
\label{JCRevR}
{\cal R}_{0 \leftarrow \tau}(\rho_\tau) = \Tr_{f}[U_{0 \leftarrow \tau} (\rho_\tau \otimes \gamma_f) U^\dagger_{0 \leftarrow \tau}]
\end{equation}
where $U_{0 \leftarrow \tau} = e^{i H_{\rm JC} \tau} = U_{0 \rightarrow \tau}^\dagger$, which is equivalent to changing the Hamiltonian $H_{\rm JC} \rightarrow -H_{\rm JC}$.

Using Theorem \ref{PureFT}, we find the symmetry of the forward and backward transition probabilities between two pure atomic states $\ket{\psi} \rightarrow \ket{\phi'}$ with the energy difference $\Delta E = E_{\phi'} - E_{\psi}  = \langle \phi' | H_a | \phi' \rangle - \langle \psi | H_a | \psi \rangle $:  
\begin{equation}
\label{QuantumThermoFT}
\frac{T ( \ket{\psi} \rightarrow \ket{\phi'} ) }{ \tilde T( | \tilde{\psi} \rangle \leftarrow | \tilde\phi' \rangle)} = \Upsilon  e^{- \beta \Delta E} ,
\end{equation}
which resembles the detailed balance condition in classical thermodynamics, with the extra factor $\Upsilon = \exp [\beta \Delta E + \log \langle \psi | e^{\beta H_a} | \psi \rangle \langle \phi' | e^{-\beta H_a} | \phi' \rangle] $. We note that the extra factor $\Upsilon$ contains the higher-order terms of the system Hamiltonians, and $\Upsilon \geq 1$. This captures the effect of coherences on the quantum detailed balance as $\Upsilon > 1$ \textit{if and only if} either the initial or final state contains energy coherences. It can be understood in the sense that fluctuations in coherences make it more difficult to achieve the reverse quantum process of the rescaled states. 

Figure \ref{PureQFTFig} compares the detailed balances regarding incoherent and coherent state-transition probabilities. The parameters for $H_{\rm JC}$ are given by $\beta = 1$, $\omega_0 =1$ and $g=0.1$.
When the initial and final states do not contain coherences in the energy-eigenstate basis, (e.g. $\ket{e} \rightarrow \ket{g}$ or $\ket{g} \rightarrow \ket{g}$, where $\ket{g}$ and $\ket{e}$ are the ground and excited atomic states) $\Upsilon$ becomes $1$. On the other hand, the transition from $\ket{\psi} = \sqrt{r} \ket{g} + \sqrt{1-r} \ket{e}$ to $\ket{\phi'} = \sqrt{1-r} \ket{g} - \sqrt{r} \ket{e}$ for $0 < r < 1$ leads to $\Upsilon > 1$ reflecting the role of coherence in the detailed balance condition. For example, by choosing $r=1/2$ to set $\Delta E = 0$, the quantum correction is given by $\Upsilon \approx 1.27$.

We point out that this can be generalized to any multi-level systems with the appropriate Hamiltonian $H$. The effect of coherence becomes significant when the temperature is low and the state has coherence between a large energy difference. In contrast, for the high temperature limit we recover the conventional fluctuation theorem. Up to the second order of $\beta$, the quantum correction is given by $\Upsilon \approx 1 + \frac{1}{2}[ \beta^2 ( {\rm Var}_{\ket{\psi}}(H) +   {\rm Var}_{\ket{\phi'}}(H) ) ]$, where ${\rm Var}_{\ket{\psi}}(H) = \bra{\psi} H^2 \ket{\psi} - \bra{\psi} H \ket{\psi}^2$. 

\subsection{Master equation approach}
Our approach can be applied to quantum Markov processes. Suppose that the dynamics of a quantum state is given by the Lindblad equation
$$
\frac{d\rho}{dt} = {\cal L} (\rho) = - \frac{i}{\hbar} [H_t, \rho] + \sum_n \left( L_n \rho L_n^\dagger - \frac{1}{2} \{ L_n^\dagger L_n, \rho\} \right),
$$
where $[A,B]= AB- BA$ is the commutator, and $\{A, B\} = AB + BA$ is the anti-commutator. When we know the full trajectory of the reference state $\gamma_t$, which evolves by the dynamics ${d \gamma_t } / { dt} = {\cal L}(\gamma_t)$ for $0 \leq t \leq \tau$, we can construct the reverse dynamics based on the Petz recovery map. This can be done by considering a channel ${\cal N}_{t\rightarrow t+dt}$ at time $t$ for an infinitesimal time interval $dt$, which can be written as
$$
{\cal N}_{t \rightarrow t+dt} = \mathbb{1} + {\cal L} dt.
$$
According to the definition of the Petz recovery map, the reverse process for this infinitesimal time interval is given by
$$
{\cal R}_{t \leftarrow t+dt} = \JJ{\gamma_{t}}{\frac{1}{2}} \circ {\cal N}_{t \rightarrow t+dt} ^\dagger \circ \JJ{\gamma_{t+dt}}{-\frac{1}{2}}.
$$
We note that this reverse dynamics can be expressed in another Lindblad superoperator such that
${\cal R}_{t \leftarrow t+dt} = \mathbb{1} + \tilde{\cal L} dt$,
where
\begin{equation}
\label{MarkovRev}
{\cal \tilde{L}} (\rho) = - \frac{i}{\hbar} [\tilde H_t, \rho] + \sum_n \left( \tilde{L}_n \rho \tilde{L}_n^\dagger - \frac{1}{2} \{ \tilde{L}_n^\dagger \tilde{L}_n, \rho \}\right).
\end{equation}
Here, the Hamiltonian and jump operators for the reverse dynamics are defined as $\tilde H_t = -\frac{1}{2} \gamma_t^{\frac{1}{2}} \left( H_t + i \hbar \gamma_t^{-\frac{1}{2}} (d \gamma_t^{\frac{1}{2}} / dt) + \frac{i\hbar}{2} \sum_n L_n^\dagger L_n \right) \gamma_t^{-\frac{1}{2}} + \rm{h.c.}$ and $\tilde L_n = \gamma_t^{\frac{1}{2}} L_n^\dagger \gamma_t^{-\frac{1}{2}}$, respectively. This reverse dynamics fully recovers the trajectory of the reference state, i.e., ${d \gamma_{\tau - \tilde{t}}} / {d\tilde{t}} =  {\cal \tilde L} (\gamma_{\tau - \tilde{t}})$, where $\tilde t$ represents the evolution time in the reverse trajectory.

This result can be compared with the reverse dynamics of a Lindblad master equation studied in Ref.~\cite{Manzano18}. First, our result provides the closed form of the reverse dynamics in terms of the valid Hamiltonian $\tilde H_t$ and $\tilde L_n$ jump operators in its master equation, while the approach in Ref.~\cite{Manzano18} results in the non-Hermitian effective Hamiltonian to describe the reverse dynamics. Second, our approach does not require any condition on the forward dynamics, thus can be applied to an arbitrary quantum Markov channel, while specific channels have been considered in Refs.~\cite{Cuetara16, Manzano18}.


We discuss an application of the reverse Markov dynamics to FTs when noise is included. Let us consider the previous examples of the JC Hamiltonian, subject to the thermal environment. The Lindbladian responsible for the noise is given by $ {\cal L}_{\rm noise}(\rho) = \Gamma ( \sigma_+ \rho \sigma_- + (1/2) \{ \sigma_- \sigma_+, \rho \}) + \Gamma e^{\beta \hbar \omega_0} ( \sigma_- \rho \sigma_+ + (1/2) \{ \sigma_+ \sigma_-, \rho \})$, where $\Gamma$ is the coupling constant with thermal environment. The dynamics of the total atom-field state is given by 
\begin{equation}
\label{JC-diss}
\frac{d \rho}{dt} = {\cal L}(\rho)=- i [H_{JC}, \rho] + {\cal L}_{\rm noise}(\rho).
\end{equation}
We note that the Gibbs state $\gamma = \gamma_a \otimes \gamma_f$ of both atom and field modes is stationary under the dynamics, i.e., ${\cal L}(\gamma) = 0$, which can thus conveniently serve as the fixed-point reference state as in the case without noise. The reverse dynamics is then given by the inversion of the Hamiltonian, $\tilde H_{JC} = - H_{JC}$, analogously to Eq.~(\ref{JCRevR}), while the noise term remains the same, $\tilde{\cal L}_{\rm noise} = {\cal L}_{\rm noise}$. If we trace out the field mode, we obtain the reverse trajectory of the atomic state. Figure~\ref{PureQFTFig} shows how the forward and backward transition probabilities between the two pure states change when the noise is added to the system's dynamics. We highlight that the detailed balance condition Eq.~(\ref{QuantumThermoFT}) is unchanged as the atomic state $\gamma_a$ remains the same for the cases with and without the noise term.

\subsection{Quantum Crooks FT}
Let us move on to derive the QFT for a mixed state transformation through the quantum channel. Suppose that $\rho = \sum_\mu p_\mu \ket{\psi_\mu} \bra{\psi_\mu}$ is transformed into ${\cal N}(\rho)  = \sum_{\nu'} p'_{\nu'} \ket{\phi'_{\nu'}} \bra{\phi'_{\nu'}}$ by the quantum channel ${\cal N}$. Here $\{ p_\mu, \ket{\psi_\mu} \}$ and $\{ p'_{\nu'}, \ket{\phi'_{\nu'}} \}$ are the eigenvalue-decompositions of $\rho$ and ${\cal N}(\rho)$ respectively, and $\ket{\psi_\mu}$ and $\ket{\phi'_{\nu'}}$ are not necessarily orthogonal to each other. The single-shot entropy change is calculated similarly to the classical channel as
$$
\delta s^{\mu \rightarrow {\nu'}}:= -\log p'_{\nu'} + \log p_\mu,
$$
based on  the von Neumann entropy $S(\rho) = -\Tr[\rho \log \rho] =  -\sum_\mu p_\mu \log p_\mu$.

The information exchange of the quantum channel is characterized with respect to the reference bases $\{ \ket i\}$ and $\{ \ket {k' }\}$. The classical and quantum channels are different as the latter could include coherences; thus, transitions between off-diagonal elements, $\ket{i}\bra{j} \xrightarrow{\cal N} \ket{k'}\bra{l'}$ should be considered as well. We define the quantum information exchange to contain all these transitions by
\begin{equation}
\delta q_{ij \rightarrow k'l'} := -\log \left[  \frac{ T_{ij \rightarrow k'l'} }{ \tilde T^*_{ij \leftarrow k'l'} } \right] = -\frac{1}{2} \log[r'_{k'} r'_{l'}] + \frac{1}{2} \log[r_i r_j],
\label{qie-r}
\end{equation}
where $T_{ij \rightarrow k'l'} := \bra{k'} {\cal N} (\ket{i} \bra{j}) \ket{l'}$ and $\tilde T_{ij \leftarrow k'l'} := \bra{i} {\cal R}_\gamma (\ket{k'} \bra{l'}) \ket{j}$. Even though the transition matrices are complex-valued, the ratio between the forward process $(T_{ij \rightarrow k'l'})$ and the backward process after taking the complex conjugate $(\tilde T_{ij \leftarrow k'l'}^*)$ is always positive. If we consider the transition between the diagonal elements $T_{ii \rightarrow k'k'}$, the quantum information exchange reduces to the classical case Eq.~(\ref{EntFluxDef}).

We note that taking the complex conjugate is necessary to establish the symmetry between the forward and backward transitions of off-diagonal elements and to consider the equivalence of the transpose operation, $\tilde T_{ij \leftarrow k'l'}^* = \tilde T_{ji \leftarrow l'k'}$. This implies that the reverse process should be redefined as ${\cal R}^*_\gamma := \Theta \circ {\cal R}_\gamma \circ \Theta'$ to include the transpose operations $\Theta (\ket{i} \bra{j}) = \ket{j}\bra{i} $ and $\Theta' (\ket{k'} \bra{l'}) = \ket{l'}\bra{k'}$, which are related to the time-reversal operation in quantum mechanics.  For example, in a harmonic oscillator system with the Hamiltonian $H= p^2/(2m) + m \omega_0^2 x^2/2$ with $[x,p]=i \hbar$, the transpose operation $\Theta$  with respect to the energy eigenstates is identical to the reflection in phase space, $p \xrightarrow{\Theta} -p$ and $x \xrightarrow{\Theta} x$, which is equivalent to the time-reversal operation.  However, ${\cal R}^*_\gamma$ is no more a CPTP map as the transpose operation does not preserve the complete-positivity condition \cite{Peres96}. This problem can be bypassed by noting that the quantum states should also be time-reversed in the reverse trajectory. Time-reversing the quantum states cancels out the effect of the transpose operations as $\Theta ( {\cal R}^*_\gamma ( \Theta'(\rho')) ) = {\cal R}_\gamma (\rho')$. Throughout this paper we will keep using the Petz recovery map and the reversed state without the time-reversal operations to preserve the CPTP property of the reverse quantum channel, as ${\cal R}_\gamma$ and $\Theta \circ {\cal R}_\gamma^* \circ \Theta'$ give the same picture for a physical state described by a density matrix.

\begin{figure}[t]
\includegraphics[width=0.8\linewidth]{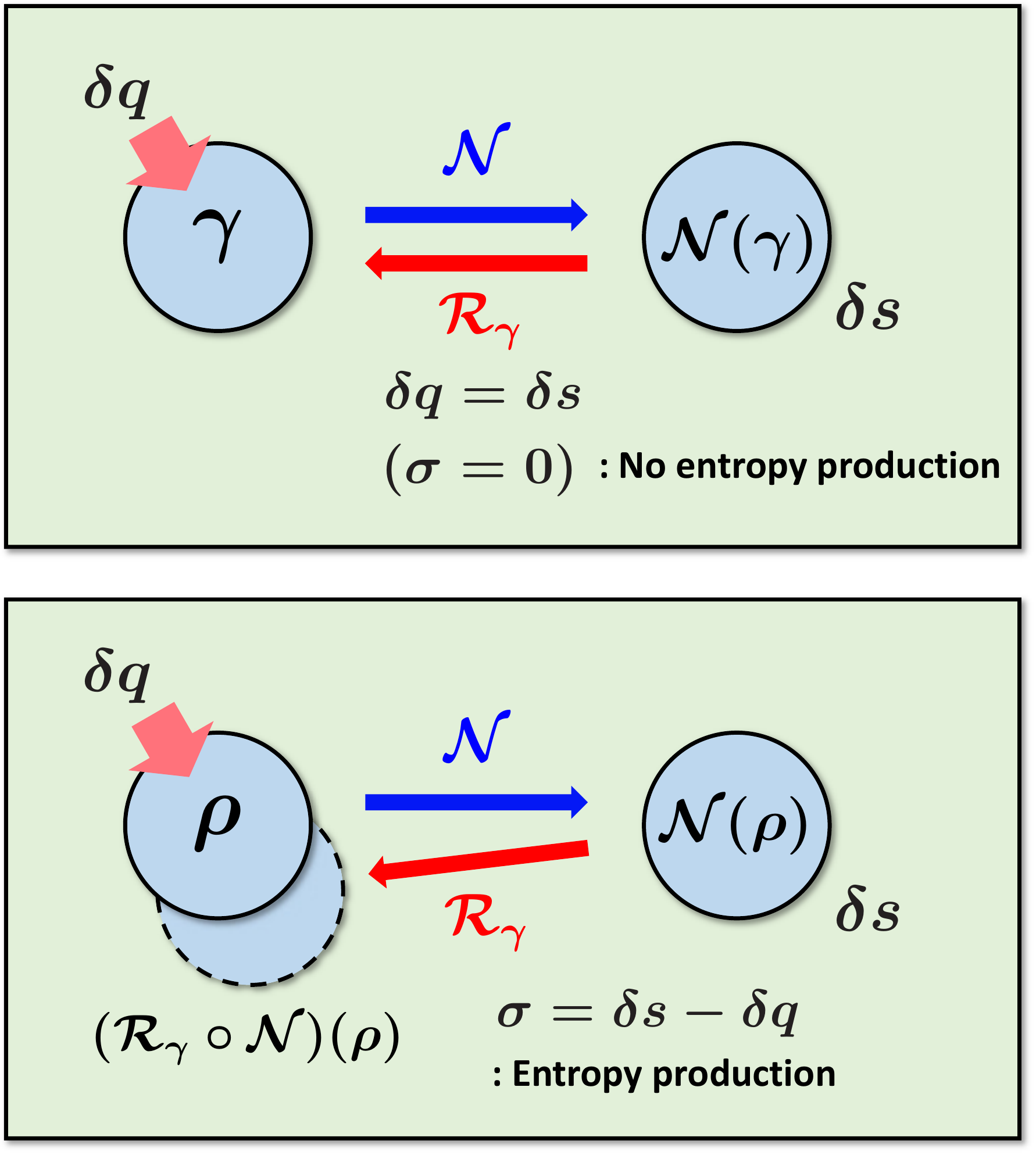}
\caption{Quantum information exchange ($\delta q$) is equal to the entropy difference ($\delta s $) for the reference state $\gamma \xrightarrow{\cal N} {\cal N}(\gamma)$ (the upper figure). When the recovery map ${\cal R}_\gamma$ does not fully recover the quantum state $\rho$, there is entropy production $\sigma = \delta s - \delta q \neq 0$ (the lower figure).}
\label{QIEFig}
\end{figure}

In a similar way to the classical channel given by Eq.~(\ref{CEntProdDef}), the entropy production for the transition $(\mu, i, j) \rightarrow (\nu', k', l')$ is defined as 
\begin{equation}
\sigma^{\mu \rightarrow {\nu'}}_{ij \rightarrow k'l'} := \delta s^{\mu \rightarrow {\nu'}} - \delta q_{ij \rightarrow k'l'}.
\label{e-production}
\end{equation}
For the reference state, the entropy production becomes zero for any transitions, which corresponds to a reversible process in thermodynamics (see Fig.~\ref{QIEFig}). 

The final step to establish the QFT is constructing the TPM distribution while keeping all the marginal distributions of $\mu, i, j, {\nu'}, k', l'$ unchanged. Although it is impossible to find a positive TPM quasi-probability distribution satisfying such a condition \cite{Perarnau-Llobet17, Lostaglio18}, we can define a complex-valued distribution for the forward process as
\begin{equation}
\label{JointProb}
P^{\mu , {\nu'}}_{ij , k'l'} := p_\mu \langle \phi'_{\nu'} | \Pi_{k'}  {\cal N} ( \Pi_i | \psi_\mu \rangle \langle \psi_\mu |  \Pi_j ) \Pi_{l'} | \phi'_{\nu'} \rangle,
\end{equation}
where $\Pi_i = \ket{i}\bra{i}$ and $\Pi_{k'} = \ket{k'}\bra{k'}$. We can prove that $P^{\mu, {\nu'}}_{ij, k'l'}$ satisfies the marginality:
\begin{equation}
\label{Marginal}
\begin{aligned}
\sum_{{\nu'}~{\rm or}~\mu} \left[ \sum_{i,j,k',l'}P^{\mu , {\nu'}}_{ij , k'l'} \right] &= p_\mu~{\rm or}~p'_{\nu'}\\
\sum_{k'~{\rm or}~i} \left[ \sum_{\mu,{\nu'},j,l'} P^{\mu , {\nu'}}_{ij , k'l'}\right] &=  {\rm Tr} \left[ \Pi_i \rho \right]~{\rm or}~{\rm Tr}[\Pi_{k'} {\cal N} (\rho)].
\end{aligned}
\end{equation}

We are now ready to derive the entropy production QFT. Using the TPM quasi-probability, the distribution of entropy production is given by
\begin{equation}
\label{EntProdDef}
P_\rightarrow(\sigma) = \sum_{\mu, i, j} \sum_{\nu',k',l'} P^{\mu , {\nu'}}_{ij , k'l'} \delta \left(\sigma - \sigma^{\mu \rightarrow {\nu'}}_{ij \rightarrow k'l'} \right),
\end{equation}
and $P_\leftarrow(\sigma)$ can be similarly defined using the recovery map ${\cal R}_\gamma$. We note that $P_\rightarrow(\sigma)$ is a real-valued function, and only the real parts of the TPM quasi-probability ${\rm Re} [P^{\mu , {\nu'}}_{ij , k'l'}]$ contribute to $P_\rightarrow(\sigma)$ as $P^{\mu , {\nu'}}_{ij , k'l'} = (P^{\mu , {\nu'}}_{ji , l'k'})^*$ and $\sigma^{\mu \rightarrow {\nu'}}_{ij \rightarrow k'l'} = \sigma^{\mu \rightarrow {\nu'}}_{ji \rightarrow l'k'}$. Analogous to the classical Crooks FT, we establish the symmetry between the forward and backward quantum transitions.
\begin{theorem} The distribution of quantum entropy production for the CPTP map ${\cal N}$ is related to its reverse process ${\cal R}_\gamma$ as: 
\label{CFT}
\begin{equation}
\label{Crooks}
\frac{P_\rightarrow(\sigma)}{P_\leftarrow (-\sigma)} = e^{\sigma}.
\end{equation}
\end{theorem}

\subsection{Reconstructing the TPM quasi-probability distribution from a two-point POVM protocol}
\label{sec:TPPOVM}
We schematically describe how to experimentally show QFT. The first task is to study whether the complex-valued TPM quasi-probability distribution $P^{\mu, \nu'}_{ij,k'l'}$ can be obtained from observable quantities. Let us consider the TPM probability 
$$
P_\rightarrow (m,m') = \Tr[ M'_{m'} {\cal N} (M_m \rho M_m^\dagger) {M'}^\dagger_{m'}],
$$ 
where $M_m$ and $M'_{m'}$ are measurement operators for the initial and final points with the measurement outcomes $m$ and $m'$, respectively. Note that $P_\rightarrow(m,m')$ is a proper probability distribution. Such protocol has recently been experimentally realized to test fluctuation theorems in the quantum regime \cite{Batalhao14, An15, Smith18, Gardas18}.

We point out that the projection measurement $\Pi_i$ and $\Pi_{k'}$ cannot directly be adopted for $M_m$ and $M'_{m'}$ as all the coherence terms vanish after the measurements are performed. In order to keep coherences, we perform POVMs at the initial and final points, where POVM elements can overlap each other. We provide a two-point POVM protocol such that the TPM quasi-probability can be obtained from the statistics of the measurement outcomes.

For our example of the JC Hamiltonian Eq.~(\ref{JC-h}), where the reference states have the eigenstates $\{\ket{g}, \ket{e} \}$, the POVMs for the first and second measurements can be given by 
$$
\begin{aligned}
M_{(\mu,1)} &= \frac{1}{\sqrt{2}}  \ket{g}\bra{g} \Pi_{\psi_\mu} \\
M_{(\mu,2)} &= \frac{1}{\sqrt{2}}  \ket{e}\bra{e} \Pi_{\psi_\mu} \\
M_{(\mu,3)} &= \frac{1}{2} (\ket{g}\bra{g} + \ket{e}\bra{e}) \Pi_{\psi_\mu} \\
M_{(\mu,4)} &= \frac{1}{2}  (\ket{g}\bra{g} + i \ket{e}\bra{e}) \Pi_{\psi_\mu}
 \end{aligned}
$$
and
$$
\begin{aligned}
M'_{(\nu',1')} &= \frac{1}{\sqrt{2}} \Pi_{\phi_{\nu'}} \ket{g}\bra{g}  \\
M'_{(\nu',2')} &= \frac{1}{\sqrt{2}} \Pi_{\phi_{\nu'}} \ket{e}\bra{e}  \\
M'_{(\nu',3')} &= \frac{1}{2} \Pi_{\phi_{\nu'}} (\ket{g}\bra{g} + \ket{e}\bra{e}) \\
M'_{(\nu',4')} &= \frac{1}{2} \Pi_{\phi_{\nu'}} (\ket{g}\bra{g} + i \ket{e}\bra{e}) .
 \end{aligned}
$$
Here, $\Pi_{\psi_\mu} = \ket{\psi_\mu}\bra{\psi_\mu}$ and $\Pi_{\phi_{\nu'}} = \ket{\phi_{\nu'}}\bra{\phi_{\nu'}}$, and the measurement outcomes can be represented as $m = (\mu, a)$ and $m' = (\nu', b')$ for all possible $(\mu, a)$ and $(\nu', b')$. In experiments, $\Pi_{\psi_\mu}$, $\Pi_{\phi_{\nu'}}$, $\ket{g}\bra{g}$, and $\ket{e}\bra{e}$ can be realized by projection measurements combined with single-qubit gates. The other POVM elements $\frac{1}{\sqrt{2}} \left( \ket{g}\bra{g} + \ket{e}\bra{e} \right) = \frac{\mathbb{1}}{\sqrt{2}}$ and $\frac{1}{\sqrt{2}} \left( \ket{g}\bra{g} + i \ket{e}\bra{e} \right) = \frac{1}{\sqrt{2}}e^{i (\pi/2) \sigma_z}$ can be obtained by performing the controlled-phase gate of $\phi=\pi/2$ along with an initial ancillary state $\ket{0}_c + \ket{1}_c$ followed by the projections onto $\ket{0}_c\bra{0}$ and $\ket{1}_c\bra{1}$. These quantum operations can be implemented in various physical systems \cite{AT1, PP1, PP2, PP3, II1, II2, SQ1, SQ2, SQ3, Exp1, Exp2} including atoms \cite{AT1}, photons \cite{PP1,PP2, PP3}, trapped ions \cite{II1,II2}, and superconducting circuits \cite{SQ1, SQ2, SQ3}.

We note that both real and complex components of the TPM quasi-probability $P^{\mu,\nu'}_{ij,k'l'}$ can be fully reconstructed as a linear combination of the two-point POVM distribution $P_\rightarrow(m,m')$. Detailed expression can be found in Appendix \ref{appx:TPPOVM}. The TPM quasi-probability for the reverse process can be obtained in a similar way by the following two-point POVMs $P_\leftarrow (m,m') = \Tr[ M_m^\dagger {\cal R}_\gamma ( {M'}^\dagger_{m'} {\cal N}(\rho) M'_{m'}  ) M_m]$. 

This protocol can be generalized for the initial and final quantum states in a $d$-dimensional Hilbert space. In this case, a set of POVMs to obtain the TPM quasi-probability can be constructed as follows:
$\{ M_m \} = \{ \frac{1}{\sqrt{d}}\Pi_i \Pi_{\psi_\mu} , \frac{1}{\sqrt{2d}}(\Pi_i +\Pi_j)\Pi_{\psi_\mu}, \frac{1}{\sqrt{2d}}(\Pi_i + i\Pi_j)\Pi_{\psi_\mu} \}$ and $\{ M'_{m'} \} = \{ \frac{1}{\sqrt{d}} \Pi_{\phi_{\nu'}} \Pi_{k'} , \frac{1}{\sqrt{2d}} \Pi_{\phi_{\nu'}} (\Pi_{k'} +\Pi_{l'}), \frac{1}{\sqrt{2d}} \Pi_{\phi_{\nu'}} (\Pi_{k'} + i \Pi_{l'}) \}$ for every possible $\mu, i, j$ and $\nu', k', l'$ satisfying $i<j$ and $k' <l'$. When the initial and final quantum states commute with their reference states, i.e., $[\rho, \gamma] = 0$ and $[{\cal N}(\rho), {\cal N}(\gamma)]=0$, the two-point POVM reduces to the conventional TPM protocol using the projectors $\{ \Pi_i \}$ and $\{ \Pi_{k'} \}$.

Once we obtain the TPM quasi-probability $P^{\mu, \nu'}_{ij,k'l'}$, the quasi-probability of the entropy production $P_\rightarrow(\sigma)$ can be reconstructed by Eq.~(\ref{EntProdDef}).

\section{Fluctuation theorems for a quantum channel inducing coherence transfer}
\label{Sec-CohChannel}
\subsection{Coherence transfer and negative quasi-probability distributions}

\begin{figure}[t]
\includegraphics[width=0.85\linewidth]{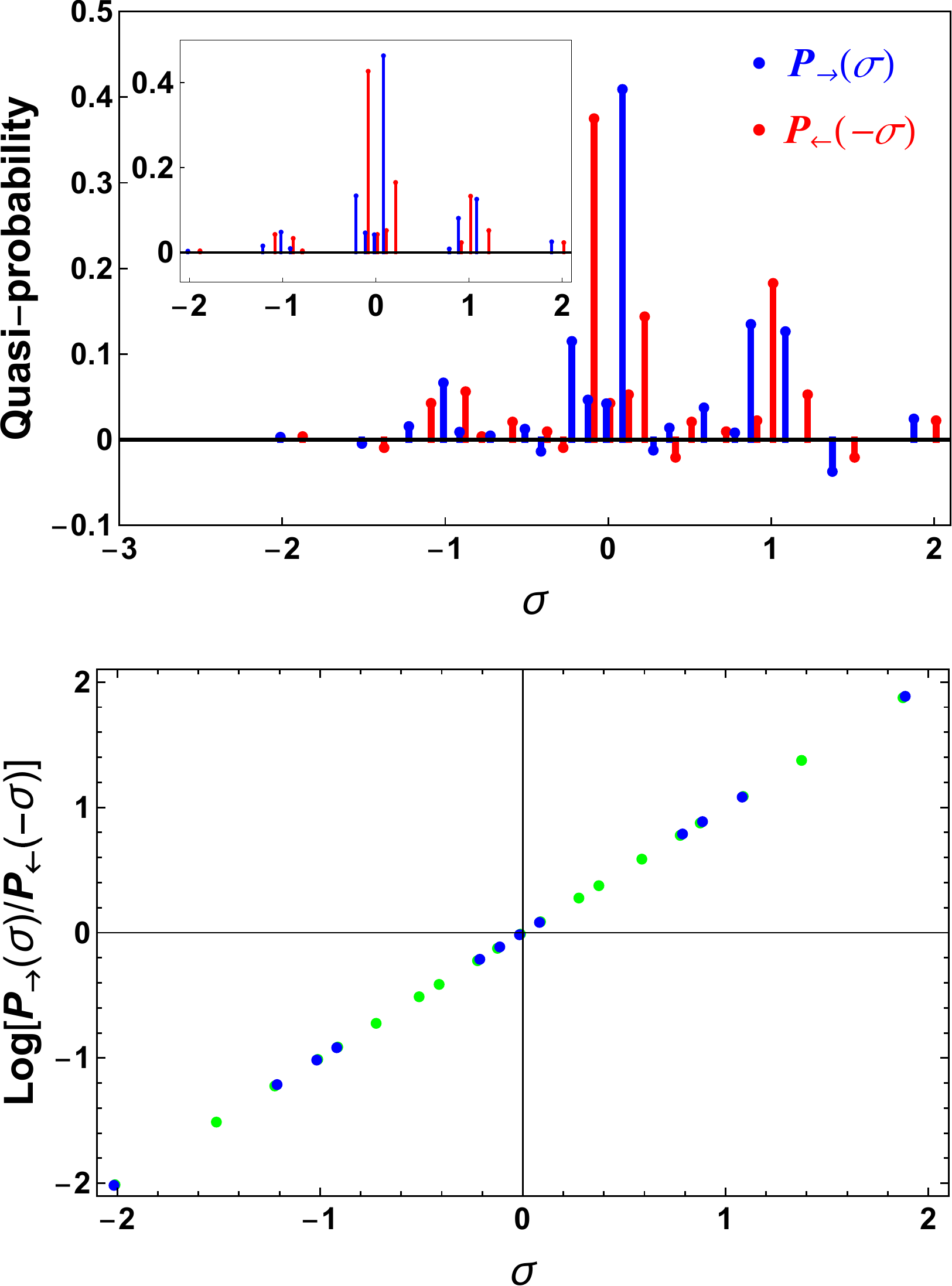}
\caption{(Upper figure) Distribution of the forward $P_\rightarrow(\sigma)$ (blue) and backward $P_\leftarrow(-\sigma)$ (red) entropy productions in the JC Hamiltonian. The same parameters for $H_{\rm JC}$ are chosen as in Fig.~\ref{PureQFTFig} with $\tau =18.66$. The initial atomic state is given by $\rho = (1/2) \ket{\psi} \bra{\psi} + \mathbb{1}/4$ with $\ket{\psi}= (\ket{g} + \ket{e})/\sqrt{2}$. When the field is initially in the incoherent thermal state $\gamma_f$, $P_\rightarrow(\sigma)$ and $P_\leftarrow(-\sigma)$ are always positive (inner figure). When the initial field is in the coherent Gibbs state $\ket{\gamma_f}$, $P_\rightarrow(\sigma)$ and $P_\leftarrow(-\sigma)$ can be negative (outer figure). (Lower figure) For both coherent (green) and incoherent (blue) cases, the entropy production satisfies the fluctuation theorem $\log \left[ P_\rightarrow(\sigma)/P_\leftarrow(-\sigma) \right] = \sigma$ in Eq.~(\ref{Crooks}).}
\label{NegEntPFig}
\end{figure}

Even though Theorem~\ref{CFT} looks remarkably similar to its classical counterpart (\ref{CFTEq}), there is a crucial difference that $P_\rightarrow (\sigma)$ can have negative values (see Fig.~\ref{NegEntPFig} and Section~\ref{Sec-Example} for detailed discussions). This stems from the fact that the real part of the TPM quasi-probability ${\rm Re} [ P^{\mu , {\nu'}}_{ij , k'l'} ]$ can be negative. The negativity in the TPM quasi-probability distribution can be understood in line with that the work quasi-probability distributions \cite{Allahverdyan14, Perarnau-Llobet17, Lostaglio18} should allow negativity to preserve the marginal distribution of work without disturbing the mean energy difference in the TPM setting. The negativity in the work distribution occurs when the quantum state and measurement operators do not commute \cite{Allahverdyan14}, and its relation to contextuality \cite{Lostaglio18} has been recently studied. More generally, we can find a connection between the TPM and Wigner quasi-probability distributions \cite{Wigner32} by noting that both preserve the marginal probabilities of non-commuting observables. In this manner, their negativities can been studied as a signature of nonclassicality \cite{Kenfack04}. Also, the deeper physical meaning of the negativity might be found based on its relationship to contextuality \cite{Spekkens08}, which we leave to future analysis.

 Let us discuss the necessary conditions to obtain the negativity in $P_\rightarrow(\sigma)$ in two different aspects: 1) coherence contained in the system and 2) coherence transfer induced by the channel. We study the condition for coherence in the quantum states first. We note that the TPM distribution is always positive when both initial and final states are diagonal with respect to the reference states, i.e, $\rho = \sum_i p_i \ket{i}\bra{i}$ and ${\cal N}(\rho) = \sum_{k'} p'_{k'} \ket{k'} \bra{k'}$. In this case, the TPM distribution is given by $p_i T (\ket{i} \rightarrow \ket{k'}) \geq 0$ and the FTs for classical channels can be retrieved. This result implies that coherence in the quantum states is a necessary condition for the negativity in  $P_\rightarrow(\sigma)$.

However, it is important to note that the initial or final quantum state containing coherence is not enough to observe the negativity in $P_\rightarrow(\sigma)$. We can have a quantum channel ${\cal N}$ which leads to positive $P_\rightarrow(\sigma)$ regardless of coherences in the initial or final state. Suppose the quantum channel ${\cal N}$ described by a set of Kraus operators $\{K_m \} $ satisfying
$
\langle k' | K_m | i \rangle = K_{ij}(\omega_m) \delta (\omega_m  + \log r'_{k'} - \log r_i ),
$
where $K_{ij}(\omega_m)$ is some complex-valued function. Through this channel, the transition $\ket{i}\bra{j} \xrightarrow{\cal N} \ket{k'}\bra{l'}$ occurs between the same mode of coherence \cite{Marvian16a} satisfying $\omega_{ij} = \omega'_{k'l'}$, where $\omega_{ij} := -\log r_i + \log r_j$ and $\omega'_{k'l'}:= -\log r'_{k'} + \log r'_{l'}$. When $\{ r_i \}$ and $\{ r'_{k'} \}$ are non-degenerate, the channel does not generate off-diagonal terms $\ket{k'}\bra{l'}$ from any diagonal terms $\ket{i}\bra{i}$. We note that this type of quantum channel has a positive distribution $P_\rightarrow(\sigma) \geq 0$ for any initial states. Therefore, the quantum channel should induce a coherence transfer between off-diagonal elements to have the negativity in $P_\rightarrow(\sigma)$. In the following sections, we will see how the coherence transfers by the quantum channel lead to a qualitatively different nature of QFTs---the imaginary entropy production.

\subsection{Rotated Petz recovery maps and imaginary quantum information exchange}
\label{CovSection}
\begin{figure}[t]
\includegraphics[width=0.8\linewidth]{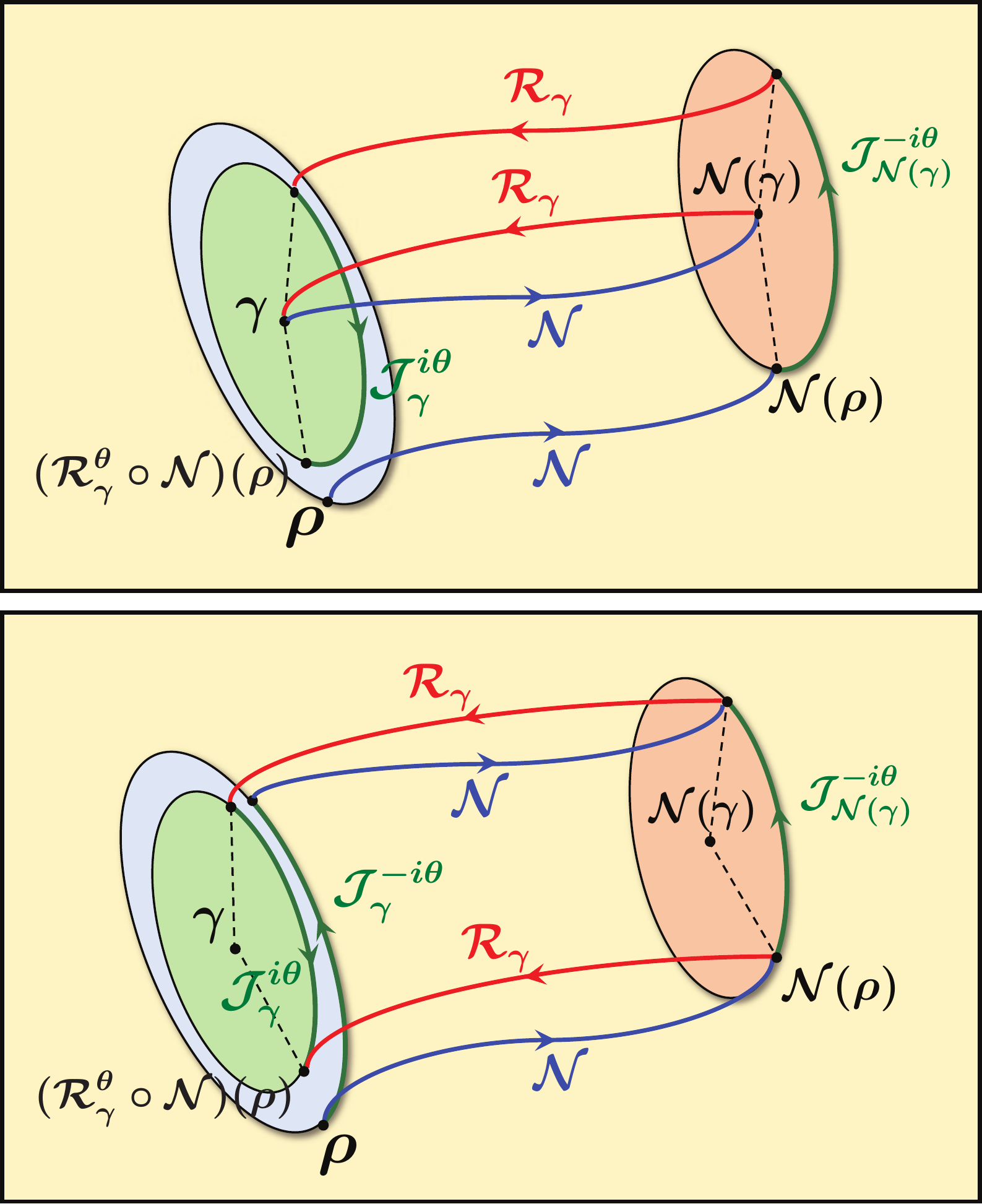}
\caption{Rotated Petz recovery maps (upper figure). The reference state $\gamma$ is fully recovered by the Petz recovery map, i.e. $({\cal R}^\theta_\gamma \circ {\cal N}) (\gamma) = \gamma$. 
For a covariant quantum channel ${\cal N}$ satisfying $\JJ{{\cal N}(\gamma)}{-i \theta} \circ {\cal N} \circ \JJ{\gamma}{i \theta} =  {\cal N}$, all the rotated Petz recovery maps are the same, i.e. ${\cal R}_\gamma^\theta = {\cal R}_\gamma$ (lower figure).}
\label{PetzFig}
\end{figure}
In order to understand the coherence transfer, it is important to note that there is an additional degree of freedom to choose in the Petz map for the full recoverability of the reference state $\gamma$ \cite{Junge15}. This additional parameter leads to a family of recovery maps called the rotated Petz recovery map (See Fig.~\ref{PetzFig}).
\begin{definition} [Rotated Petz recovery map]
For a given reference state $\gamma$ and CPTP map ${\cal N}$, the Petz recovery map can be generalized to the following form of the rotated recovery map \cite{Junge15}
\begin{equation}
\label{RotPetzDef}
{\cal R}_\gamma^{\theta}(\chi) := \left( \JJ{\gamma}{\frac{1}{2}+i \theta} \circ {\cal N}^\dagger \circ \JJ{{\cal N}(\gamma)}{-\frac{1}{2} - i \theta} \right) (\chi).
\end{equation}
Every rotated Petz recovery map fully recovers the reference state, i.e. ${\cal R}_\gamma^\theta ({\cal N}(\gamma)) = \gamma$ for every $\theta$.
\end{definition}
The ordinary Petz recovery map ${\cal R}_\gamma$ is the special case of the rotated Petz map with $\theta=0$. Now, we calculate the ratio between $P_\rightarrow$ and $P_\leftarrow$ as in Eq.~(\ref{qie-r}) with the generalized Petz map, where the reverse transition is given by $ \tilde T^{\theta}_{ij \leftarrow k'l'} = \langle i | {\cal R}_\gamma^\theta ( |k' \rangle \langle l'|) | j \rangle$. In this case, the ratio between the forward and backward transition acquires an additional phase factor as 
\begin{equation}
 \label{qie-rQ}
\frac{T_{ij \rightarrow k'l'}} {(\tilde T^{\theta}_{ij \leftarrow k'l'})^*} = e^{ - \delta q_{ij \rightarrow k'l'} + i  \theta (\omega_{ij} - \omega'_{k'l'})}.
\end{equation}
The involvement of the phase is purely due to coherences as it vanishes for the transition between diagonal elements. The rotation in the Petz recovery map modifies the quantum information exchange by adding the imaginary term:
$$(\delta q_I)_{ij \rightarrow k'l'} := \frac{\omega'_{k'l'} -\omega_{ij}  }{2}=- \frac{1}{2} \log  \left( \frac{r'_{k'}}{r'_{l'}} \right)+ \frac{1}{2} \log  \left( \frac{r_i}{r_j} \right),$$
while keeping the real part as in Eq.~(\ref{qie-r}). We choose the factor $(1/2)$ in $\delta q_I$ to be consistent with the real part of $\delta q$. The single-shot entropy production then becomes complex-valued as 
\begin{equation}
\label{e-productionQ}
\sigma^{\mu \rightarrow {\nu'}}_{ij \rightarrow k'l'} = \delta s^{\mu \rightarrow {\nu'}} - [(\delta q_R)_{ij \rightarrow k'l'} + i (\delta q_I)_{ij \rightarrow k'l'}].
\end{equation}
The quasi-probability distribution of entropy production for the forward process $P_\rightarrow(\sigma) = P_\rightarrow(\sigma_R + i \sigma_I)$ can be defined in a similar way to Eq.~(\ref{EntProdDef}), as well as for the backward process $P_\leftarrow^\theta(\sigma)$ by using the reverse process ${\cal R}_\gamma^\theta$. Analogous to the case of real-valued entropy production, $P_\rightarrow(\sigma)$ and $P_\leftarrow^\theta(\sigma)$ can be reconstructed from the two-point POVM introduced in Section~\ref{sec:TPPOVM}.

Now we present the main result of the paper, which fully reflects the involvement of quantum coherences in the channel or in the system. 
\begin{theorem}[Generalized QFT] The quasi-probability distribution of quantum entropy production $P_\rightarrow(\sigma)$ for a CPTP map ${\cal N}$ is related to its reverse process ${\cal R}_\gamma^\theta$ by
\label{RotatedCFT}
\begin{equation}
\label{RotatedCrooks}
\frac{P_\rightarrow(\sigma)}{P^{\theta}_\leftarrow (-\sigma^*)} = e^{\sigma_R - 2 i \theta \sigma_I}.
\end{equation}
\end{theorem}
Note that Theorem~\ref{CFT} can be deduced from the generalized QFT in Eq.~(\ref{RotatedCrooks}) by taking $\theta = 0$ and summing over all the imaginary entropy productions $\sigma_I$. The imaginary part of Eq.~(\ref{RotatedCrooks}) only comes from the coherence transfer between off-diagonal elements, and it arises when coherences are involved in both system and channel. In the following sections, relationships between the imaginary entropy production and the covariance property of the quantum channel are discussed. We will also see that the imaginary parts play an important role in recovering other fluctuation theorems and the second law of thermodynamics for a quantum channel. 

\subsection{Two-level atom interacting with the coherent and incoherent heat baths}
\label{Sec-Example}
As a specific case study, we recall the previous example of the JC Hamiltonian in Eq.~(\ref{JC-h}). We first show a simple case of the QFT for an incoherent thermal channel, which does not show negativities. From the definition of the entropy production Eq.~(\ref{e-productionQ}), we find $\sigma^{\mu \rightarrow {\nu'}}_{ij \rightarrow k'l'} = -\log p'_{\nu'} + \log p_\mu - \beta( E_{k'} - E_i + E_{l'} - E_j)/2 - i \beta( E_{k'} - E_{l'} - E_i  + E_j  )/2$, where $E_i$ are the energy eigenstates of the atomic Hamiltonian $H_a$. Note that $E_{k'} - E_{l'} = E_i - E_j$ should hold when the two-level atom is interacting with the thermal field. Using this condition, the quantum entropy production for the transition $(\mu, i, j) \rightarrow ({\nu'}, k', l')$ is simplified as $\sigma^{\mu \rightarrow {\nu'}}_{ij \rightarrow k'l'} = -\log p'_{\nu'} + \log p_\mu - \beta( E_{k'} - E_i )$, where $E_{k'} - E_i (= E_{l'} - E_j)$ is the energy exchange, or heat flow, from the field to the atomic system. Note that there is no imaginary part that appears in the entropy production, and the entropy production probability is always positive (see Fig.~\ref{NegEntPFig}).
\begin{figure}[t]
\includegraphics[width=0.95\linewidth]{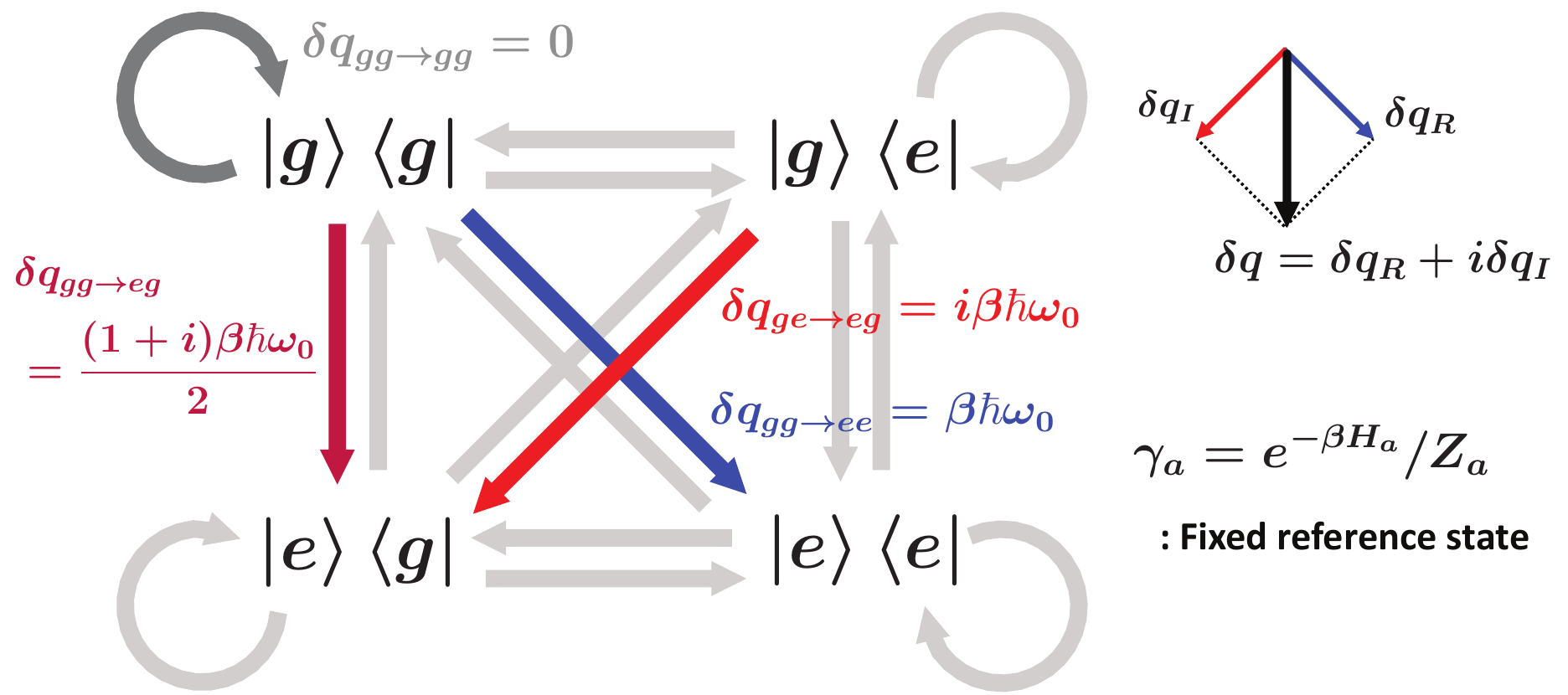}
\caption{Quantum information exchange for the two-level atom with the fixed reference state $\gamma_a = e^{-\beta H_a}/Z_a$. All possible transitions are described by the arrows, and some transformations are selected to show the values of quantum information exchange. This can be generalized to a higher dimensional system.}
\label{QIEDiagram}
\end{figure}
\begin{figure*}[t]
\includegraphics[width=.95\linewidth]{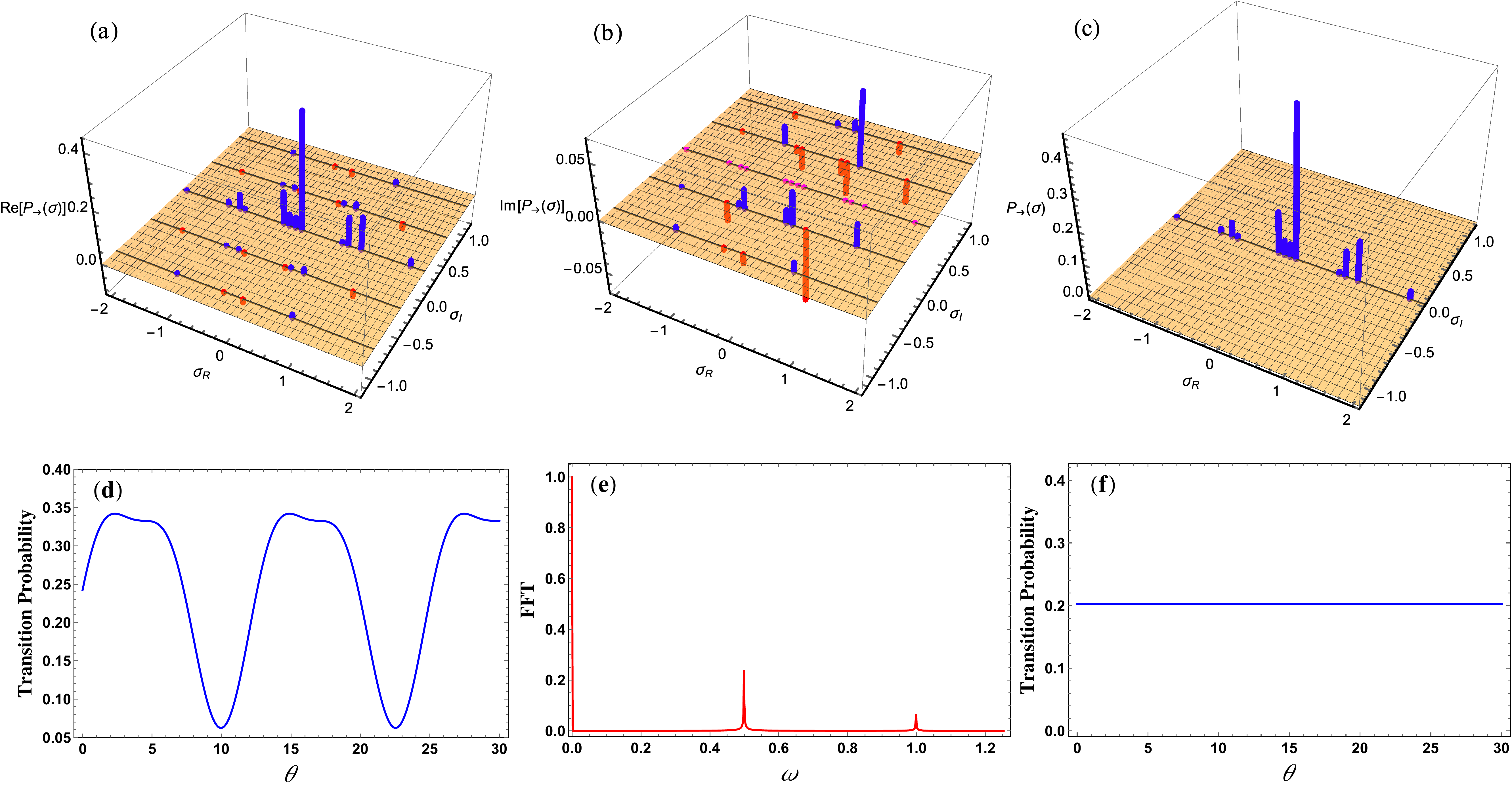}
\caption{(a) Real and (b) imaginary parts of $P_\rightarrow(\sigma_R + i \sigma_I)$ for the JC model interacting with the coherent Gibbs state $\ket{\gamma_f}$. The initial state and the parameters are the same as in Fig.~\ref{NegEntPFig}. The red points indicate the negativities. (d) The transition probability between the rotated eigenstates $T(e^{-i \beta H_a \theta/2} \ket{\psi_\mu} \rightarrow e^{-i \beta H_a \theta/2}\ket{\phi'_{\nu'}}) $ versus $\theta$ and (e) its Fourier transform with normalization. The peaks indicate the points where the imaginary entropy productions occur (i.e., $ \sigma_I = 0, \pm \beta \hbar \omega_0/2, \pm \beta \hbar \omega_0 $). (f) When the initial field is in the Gibbs state, the transition probability does not depend on the rotation, and (c) no imaginary entropy production occurs.}
\label{ImaginaryQIEFig}
\end{figure*}
When the channel induces a coherence transfer between off-diagonal elements, the situation can be different. As an example, we assume that the field is initially in a coherent Gibbs state $\ket{\gamma_f} \propto \sum_{n=0}^\infty \exp [-n \beta \hbar \omega_0/2 ] \ket{n} $ before normalization. Then the channel in Eq.~(\ref{JCChannel}) becomes
\begin{equation}
\label{JCQChannel}
{\cal N}^{\rm coh}_{0 \rightarrow \tau}(\rho_0) = \Tr_{f} [U_{0 \rightarrow \tau} (\rho_0 \otimes \ket{\gamma_f}\bra{\gamma_f}) U^\dagger_{0 \rightarrow \tau}].
\end{equation}
We choose the interaction time $\tau \neq 0$, where the thermal equilibrium state of the atom $\gamma_a$ returns to the initial thermal equilibrium, i.e. ${\cal N}_{0\rightarrow \tau}^{\rm coh} (\gamma_a) = \gamma_a$. This allows us to keep $\gamma_a$ as the fixed reference state as for the thermal channel.

Although the reference state $\gamma_a$ is unchanged by the channel, coherence transfers occur from the coherent bath to the system as the condition $E_{k'} - E_{l'} = E_i - E_j$ no longer holds. We characterize the transitions between off-diagonal elements by using the complex-valued quantum information exchange (see Fig.~\ref{QIEDiagram}). For example, a transition $\ket{g} \bra{g} \xrightarrow {{\cal N}^{\rm coh}_{0 \rightarrow \tau}} \ket{e} \bra{g}$ gives the complex quantum information exchange of $\delta q_{gg \rightarrow eg} = (1+i) \beta\hbar \omega_0 /2 $, while another transition $\ket{g} \bra{e} \xrightarrow {{\cal N}^{\rm coh}_{0 \rightarrow \tau}} \ket{e} \bra{g}$ gives $\delta q_{ge \rightarrow eg} = i \beta\hbar \omega_0 $. Consequently, the quantum entropy production should be complex-valued for the cases inducing coherence transfers. Figure~\ref{ImaginaryQIEFig} shows the points of non-vanishing imaginary entropy production when the two-level atom is interacting with the coherent bath.

\subsection{Covariant quantum channel and symmetry breaking}
We discuss the physical meaning of the rotation in the Petz recovery map in relation to the covariance property of the quantum channel. We note that the rotation part in Eq.~(\ref{RotPetzDef}) can be written as $\JJ{\gamma}{i \theta} (\rho) = U(\theta)  \rho U^\dagger(\theta)$ with $U(\theta) = e^{i \theta \log \gamma}$, which is the form of the unitary group transformation given by the generator $\log \gamma$. Such transformation can be regarded as translation in time, space, or rotation, when the group generator is given by a Hamiltonian, momentum operator, or angular momentum operator, respectively. For example, when the fixed reference state is given in the form of $\gamma \propto e^{- \beta H}$ for a Hamiltonian $H$, the rotation $\JJ{\gamma}{i  \theta } (\rho) = e^{-i \beta H  \theta } \rho e^{i \beta H  \theta }$ corresponds to the time-translation operation.

The physical symmetry of the quantum channel can be studied by comparing the effects of the group transformations before and after passing the channel. When the quantum channel ${\cal N}$ satisfies the following symmetry condition with respect to the two group transformations $U(\theta)$ and $V(\theta)$:
\begin{equation}
\label{CovDef}
{\cal N} (U(\theta) \rho U^\dagger(\theta))= V(\theta) {\cal N} (\rho) V^\dagger(\theta)
\end{equation}
for all $\theta$, the channel is called covariant. In the previous example of time-translation, if the quantum channel ${\cal N}$ satisfies ${\cal N}(e^{- i \beta H \theta} \rho e^{i \beta H \theta}) = e^{- i \beta H \theta} {\cal N}( \rho )e^{i \beta H \theta}$, we say the channel is covariant with respect to the group transformations $U(\theta)$ and $V(\theta)$, where $U(\theta) = e^{-i \beta H \theta} = V(\theta)$. We also note that the covariance properties of quantum channels have been studied in the context of the resource theory of asymmetry \cite{Skotiniotis12, Marvian13, Marvian14},  quantum thermodynamics \cite{Brandao13, Gour17, LostaglioX, Kwon18}, and quantum error correction \cite{Faist19, Woods19}.

We demonstrate that the imaginary quantum information exchange is directly related to the covariance property of the channel with respect to the group transformations $U(\theta) = e^{i \theta \log \gamma}$ and $V(\theta) = e^{i \theta \log {\cal N}(\gamma)}$. We find that the imaginary part of the information exchange vanishes when the quantum channel is covariant with respect to $U(\theta)$ and $V(\theta)$. This condition can be equivalently expressed as
$$
\JJ{{\cal N}(\gamma)}{-i \theta} \circ {\cal N} \circ \JJ{\gamma}{i \theta}= {\cal N}.
$$
Conversely, the non-vanishing imaginary quantum entropy production implies that the quantum channel does not have the covariance property. We summarize the relationship between imaginary entropy production and covariance of the quantum channel as the following theorem:
\begin{theorem}[QFT for a covariant quantum channel]
\label{CovThm} For a covariant quantum channel ${\cal N}$ with respect to the reference state $\gamma$, every rotated Petz recovery map reduces into the ordinary Petz map $({\cal R}^ \theta _\gamma = {\cal R}_\gamma)$, and the quantum entropy production does not have imaginary values. 
\end{theorem}
Figure~\ref{PetzFig} provides an illustrative description of the covariant quantum channel and how every rotated Petz recovery map is reduced into the ordinary Petz recovery map. A trivial example of covariant processes is a fully decohered quantum channel given by
$$
{\cal N}_{\rm incoh} = {\cal D}_{ {\cal N} (\gamma)} \circ {\cal M} \circ {\cal D}_\gamma,
$$
where ${\cal M}$ can be any CPTP map, and ${\cal D}_\gamma(\rho) = \displaystyle \lim_{\Delta \theta  \rightarrow \infty} \frac{1}{\Delta  \theta } \int_{-\Delta  \theta /2}^{\Delta  \theta /2} d \theta  \JJ{\gamma}{i \theta } (\rho)$ is a decohering operation with respect to the eigenstates of $\gamma$.
Nontrivial covariant operations include thermal operations obeying energy-conservation law, which we have discussed previously using the JC Hamiltonian interacting with the incoherent bath.

On the other hand, the time-translational symmetry breaking occurs for the channel ${\cal N}_{0\rightarrow \tau}^{\rm coh}$ when the atom is interacting with the coherent Gibbs state in the JC Hamiltonian, which can be inferred from the imaginary entropy production (see Fig.~\ref{ImaginaryQIEFig}). The imaginary entropy production not only indicates the broken symmetry of the quantum channel but also provides additional information about how it reacts by the group transformations $U(\theta) = e^{i \theta \log \gamma}$ and $V(\theta) = e^{i \theta \log {\cal N}(\gamma)}$ in the frequency domain. We note that the value of imaginary entropy production $\sigma_I$ having non-vanishing $P_\rightarrow(\sigma_R + i \sigma_I)$ should match the peaks in the Fourier transform of the transition probability between the rotated eigenstates $T(U(\theta/2) \ket{\psi_\mu} \rightarrow V(\theta/2) \ket{\phi'_{\nu'}})$ (see Fig.~\ref{ImaginaryQIEFig}). If the channel is covariant, the transition probability does not depend on $\theta$, which is consistent with the fact that there is no imaginary entropy production.

In the following section, we demonstrate that the imaginary part of quantum entropy production also plays a crucial role in the derivation of the second law inequality from the general form of the integral QFT.

\section{Generalized second law for a quantum channel}
\label{Sec-2ndLaw}
\subsection{Integral QFT and quantum data processing inequality}
\label{IntQFTSection}
When the initial state $\rho$ has the same rank as the reference state $\gamma$, the following equality holds for every $ \theta $
\begin{equation}
\label{IntQFT}
\left< e^{-\sigma_R + i \theta  \sigma_I} \right> = 1,
\end{equation}
where $\langle \cdot \rangle$ denotes averaging over the TPM distribution. By taking $\theta =0$, we obtain $\langle e^{-\sigma_R} \rangle = \sum_{\sigma_R} P_\rightarrow (\sigma_R) e^{-\sigma_R} = 1$, which resembles the classical integral FT \cite{Seifert05}. Despite the negativities in $P_\rightarrow(\sigma)$, we verify that the examples discussed in the JC Hamiltonian in Fig.~\ref{ImaginaryQIEFig} satisfy the integral QFT given by Eq.~(\ref{IntQFT}) for both cases of coherent and incoherent baths.

However, due to the complex-valued nature of both the transition amplitude and the quantum entropy production, it is impossible to apply Jensen's inequality to obtain the inequality for the first-order moment of Eq.~(\ref{IntQFT}). This gives rise to the highly-nontrivial question of whether the QFT involving complex values implies the physically meaningful second law of thermodynamics. Nevertheless, we show that the generalized second law can be obtained by taking into account of both real and imaginary parts of quantum entropy production as follows:
\begin{theorem} [The second law for a quantum channel] The integral QFT for a quantum channel ${\cal N}$ is equivalent to the following equality 
\label{2ndLawThm}
\begin{equation}
\label{IntQFT2}
\Tr \left[ \left( {\cal J}_{{\cal N}(\rho)}^{\frac{1-i\theta }{2}} \circ {\cal J}_{{\cal N}(\gamma)}^{-\frac{1-i\theta }{2}}\circ {\cal N} \circ {\cal J}_\gamma^{\frac{1-i\theta }{2}}  \circ {\cal J}_\rho^{-\frac{1-i\theta }{2}}\right) (\rho) \right] = \kappa_\theta
\end{equation}
for any real value of $\theta$. Here, $\kappa_\theta = \Tr[ \Pi_\rho ({\cal R}_\gamma^{\theta/2} \circ {\cal N})(\rho))]$ is given by the projection  $\Pi_\rho$ onto the support of $\rho$. $\kappa_\theta =1$ when the initial state $\rho$ has the same rank with $\gamma$. From this equality condition, we can obtain a generalized second law:
\begin{equation}
\label{2ndlaw}
\left< \sigma \right> = S\left(\rho || \gamma \right) - S\left({\cal N} (\rho) || {\cal N} (\gamma) \right)  \geq 0
\end{equation}
for the quantum channel ${\cal N}$. 
\end{theorem}
It is important to note that the expectation value of the real part of entropy production is equal to the quantum relative entropy difference, and Eq.~(\ref{2ndlaw}) is known as the quantum data processing inequality \cite{Junge15}. The expectation value of the imaginary part of entropy production vanishes, i.e., $\left< \sigma_I \right> = 0$. Non-decreasing of the first-order moment of entropy production can be understood as the second law for a quantum channel from the generalized QFT similar to the relationship between the second law of thermodynamics and the classical FT. The mean entropy production $\langle \sigma \rangle$ has the physical meaning of average information loss through the noisy quantum channel as $\langle \sigma \rangle = \langle \delta s \rangle - \langle \delta q \rangle \geq 0$ implies that the system gains an additional amount of uncertainty (i.e., entropy) compared with the prediction from the reference state. The expectation values of higher-order momenta should obey
$$
\sum_{k=1}^\infty \frac{\left< (- \sigma_R + i\theta \sigma_I)^k \right>} {k!} = 0
$$
for any value of $\theta$ from Eq.~(\ref{2ndlaw}), so that the fluctuations of the higher-order moments can be inferred from the expectation values of lower-order moments.

By recalling the relationship between the second law and reversibility of thermodynamic processes, it is natural to ask whether $\langle \sigma \rangle = 0$ implies the perfect reversibility of the quantum state using the recovery map. In fact, this has been proven to be true by the stronger version of the quantum data processing inequality  \cite{Junge15}
\begin{equation}
\label{IrrvIneq}
\begin{aligned}
&S(\rho || \gamma) - S({\cal N} (\rho) || {\cal N} (\gamma)) \\
&\qquad\qquad \geq - \int_{-\infty}^\infty d\theta  g_0(\theta ) \log \left[ {\cal F}(\rho, ({\cal R}_\gamma^{\theta /2} \circ {\cal N})(\rho)) \right]\\
&\qquad\qquad \geq - \log \left[ {\cal F}(\rho, (\bar{{\cal R}}_\gamma \circ {\cal N})(\rho)) \right],
\end{aligned}
\end{equation}
where ${\cal F}(\rho,\tau) = || \sqrt{\rho} \sqrt{\tau} ||_1^2$ is the quantum fidelity, and $\bar{{\cal R}}_\gamma(\rho) := \int_{-\infty}^\infty d\theta g_0(\theta) {\cal R}_\gamma^{\theta /2} (\rho)$ with $ g_0(\theta ) = \frac{\pi/2}{\cosh(\pi \theta) + 1}$.  We prove this inequality in an alternative way in the Appendix. Using Eq.~(\ref{IrrvIneq}), we see that if $\langle \sigma \rangle =0$, ${\cal F}(\rho, ({\cal R}_\gamma^{\theta /2} \circ {\cal N})(\rho))$ should be $1$ for every $\theta$ which implies that $\rho$ is fully recovered by every rotated Petz recovery map. Conversely, by applying the quantum data processing inequality to the channel ${\cal R}_\gamma^\theta$ starting from ${\cal N}(\rho)$, we can see that $\langle \sigma \rangle = 0$ if $({\cal R}_\gamma^\theta \circ {\cal N}) (\rho) = \rho$. We summarize the necessary and sufficient conditions for the reversibility of the quantum channel as follows:

\begin{theorem}[Reversibility condition for a quantum channel \cite{Junge15}]
\label{CovRevThm}
For a quantum channel ${\cal N}$ with respect to the reference state $\gamma$, a transformation $\rho \xrightarrow{{\cal N}} {\cal N} (\rho)$ is fully reversible by every Petz recovery map ${\cal R}_\gamma^\theta$, {\it if and only if} the mean entropy production is zero, i.e. $\langle \sigma \rangle =0$. Also, there exists a convex sum of the rotated Petz recovery maps ${\cal \bar{R}}_\gamma$ satisfying
$$ \langle \sigma \rangle \geq -\log {\cal F}(\rho, ({\cal \bar{R}}_\gamma \circ {\cal N})(\rho)).$$
\end{theorem}

In order to relate $\langle \sigma \rangle$ to the second law of thermodynamics, we consider a thermodynamic process including the Hamiltonian change of the system and bath. Suppose that the Hamiltonians of the system and bath are given by $H_S$ and $H_B$ initially, and change into $H_S'$ and $H_B'$, respectively. The thermodynamic channel for the system interacting with the thermal bath $\gamma_B$ can have the following form:
$$
{\cal N}_{\rm th} (\rho_S) = {\rm Tr}_B [U (\rho_S \otimes \gamma_B) U^\dagger].
$$
Here, $U$ is a unitary interaction between the system and bath satisfying $U(H_S + H_B)U^\dagger = H'_S + H'_B$ to obey the energy conservation law $\langle H_S + H_B \rangle_{\rho_{SB}} = \langle H'_S + H'_B \rangle_{U \rho_{SB}U^\dagger}$ for any system-bath state $\rho_{SB}$. The two-level atom interacting with the incoherent bath by the JC Hamiltonian is the specific example of such thermodynamic channel, in which the Hamiltonians of the system and bath do not change.

In quantum thermodynamics, it has been studied that the free energy can be generalized into the form of $F(\rho) = \Tr [ \rho H]  -  k_B T S(\rho)$ to describe the second law regarding thermodynamic processes in the quantum regime \cite{Lostaglio15, Cwiklinski15, Brandao15}. This stems from an approach \cite{Bergmann55} to define non-equilibrium free energy by including information theoretic quantities. By applying the QFT to the channel ${\cal N}_{\rm th}$, we can find the FT of the free energy loss through quantum thermodynamic processes. If we take the reference state to be the Gibbs state $\gamma_S$ for the initial Hamiltonian of the system $H_S$, the thermodynamic channel ${\cal N}_{\rm th}$ maps this state into ${\cal N}_{\rm th} (\gamma_S)= \gamma'_S$ which is another Gibbs state for the final Hamiltonian $H_S'$. We note that the single-shot free energy difference can be written in terms of the entropy production as $(\delta F)^{\mu \rightarrow {\nu'}}_{ij \rightarrow k'l'} = E'_{k'} - E_i - k_B T \delta s^{\mu \rightarrow {\nu'}} = - k_B T\sigma^{\mu \rightarrow {\nu'}}_{ij \rightarrow k'l'} +\Delta F_{\rm eq}$, where $\Delta F_{\rm eq} = k_B T \log (Z_S/Z_S')$ is the equilibrium free energy difference. The imaginary entropy production vanishes due to the condition $E'_{k'} - E'_{l'} = E_i - E_j$ \cite{LostaglioX}. Then the integral QFT leads to the balanced equality relation for the quantum free energy difference for any non-equilibrium initial state with full rank as
$$
\langle e^{ \beta \delta F }\rangle =  e^{\beta \Delta F_{\rm eq}}.
$$
The second law of quantum thermodynamics can be obtained from Theorem~\ref{2ndLawThm} as
$$
\Delta F := \langle \delta F \rangle \leq \Delta F_{\rm eq}.
$$
The physical meaning of the above inequality is that the mean dissipated free energy of non-equilibrium states $\Delta F_{\rm eq} - \Delta F$ should always be greater than or equal to zero, which can be deduced from its fluctuation theorem $\langle e^{ - \beta (\Delta F_{\rm eq} - \delta F) } \rangle = 1$. Furthermore, the following reversibility condition can be obtained from Theorem \ref{CovRevThm}:
\begin{corollary}[Recovery of thermodynamic channels]
\label{ThermoRev}
For the thermodynamic channel ${\cal N}_{\rm th} (\rho) = {\rm Tr}_B [U (\rho \otimes \gamma_B) U^\dagger]$, the reverse thermodynamic channel given by the Petz recovery map ${\cal R}_\gamma ( \rho' ) = {\rm Tr}_B [U^\dagger ( \rho' \otimes \gamma'_B) U]$ fully recovers the initial quantum state \textit{if and only if} $\Delta F = \Delta F_{\rm eq}$. More precisely, the recovery fidelity of the reverse process is lower bounded by the free energy difference as
$$
{\cal F}(\rho, ({\cal R}_\gamma \circ {\cal N}_{\rm th})(\rho)) \geq e^{\beta ( \Delta F - \Delta F_{\rm eq}) }.
$$
\end{corollary}
In generalized QFT, however, the fluctuating quantity is not necessarily energetic values, but can be characterized in various physical contexts depending on the choice of reference states. In the following sections, we will see how the fluctuations of quantum information quantities in the resource theory of asymmetry \cite{Marvian13, Marvian14} and entanglement can be understood under the framework of the QFT.

\subsection{Asymmetry fluctuation in covariant channels}
\label{subsec-asymm}
Let us consider a covariant channel ${\cal N}_{\rm cov}$ with respect to the generator $L$ satisfying
\begin{equation}
\label{CovChannelDef2}
{\cal N}_{\rm cov} ( e^{- i L t} \rho e^{i L t} ) = e^{- i L t} {\cal N}_{\rm cov} (\rho) e^{i L t} 
\end{equation}
as we discussed in Section \ref{CovSection} by taking the group transformations $U(t) = e^{-iLt} = V(t)$. In the viewpoint of quantum resource theory \cite{ResourceTheoryRef}, asymmetry contained in a quantum state serves as a resource for the reference frame alignment \cite{Skotiniotis12} and quantum clocks \cite{Marvian16, Gour17, Kwon18}. The degree of asymmetry can be quantified by the relative entropy \cite{Marvian16a}
$$
C (\rho) := S(\rho || {\cal D}(\rho)),
$$
where $\displaystyle {\cal D}(\rho) = \lim_{\Delta t \rightarrow \infty} \frac{1}{\Delta t} \int_{-\Delta t/2}^{\Delta t/2} dt e^{-iL t }\rho e^{iLt} = {\cal D}_{\exp[L]} (\rho)$. When $L = \lambda_i \ket{i} \bra{i}$ does not have degenerate eigenvalues, ${\cal D}(\rho) = \sum_i \langle i | \rho | i \rangle \ket{i} \bra{i}$ becomes the diagonalized state of $\rho$. $C (\rho)$ does not increase by any covariant operations satisfying Eq.~(\ref{CovChannelDef2}), i.e., $C(\rho) \geq C({\cal N}_{\rm cov}(\rho))$, known as the monotonicity of the relative entropy of asymmetry.

The decreased amount of asymmetry through the covariant channel is due to the dissipation of asymmetry, or the asymmetry loss, which is quantified by
$$
\Delta C := C({\cal N}_{\rm cov}(\rho)) - C(\rho) \leq 0.
$$
In order to investigate the fluctuation of dissipated asymmetry based on the QFT, we choose the reference state ${\cal D} (\rho) = \sum_i r_i \ket{i} \bra{i}$. Then by using the property of the covariant channel, we see that ${\cal N}_{\rm cov} ({\cal D} (\rho)) = {\cal D}({\cal N}_{\rm cov} ( \rho)) = \sum_{k'} r'_{k'} \ket{k'}\bra{k'}$, where both $\ket{i}$ and $\ket{k'}$ are the eigenstates of $L$. When $L$ is nondegenerate, $r_i = \langle i | \rho | i \rangle$ and $r'_{k'} = \langle k' | {\cal N} (\rho) | k' \rangle$. We note that by using these reference states, the single-shot coherence loss $\delta C$ for the transition $(\mu, i, j) \rightarrow ({\nu'}, k', l')$ can be written as $\delta C^{\mu \rightarrow {\nu'}}_{ij \rightarrow k'l'} = -\sigma^{\mu \rightarrow {\nu'}}_{ij \rightarrow k'l'}$ by noting that $\Delta C = \sum_{\mu, {\nu'}} \sum_{i,j,k',l'} P^{\mu, {\nu'}}_{ij, kl} \delta C^{\mu \rightarrow {\nu'}}_{ij \rightarrow k'l'} = - \langle \sigma \rangle $ based on the TPM distribution and the single-shot quantum entropy production. The asymmetry loss then obeys the following  fluctuation relation
$$
\left< e^{ \delta C_R - i \theta \delta C_I} \right> = \kappa_\theta
$$
for any $\theta$, from the integral QFT. Note that $\kappa_\theta =1$ when the initial state $\rho$ is full rank. \textit{This fluctuation relation provides information on the statistics of the dissipated asymmetry in more detail compared to the mean asymmetry loss given by the monotonicity of the asymmetry measure $\Delta C =  \langle \delta C \rangle \leq 0$.} The inequality condition for the mean asymmetry loss is the consequence of the former equality condition as seen in Theorem~\ref{2ndLawThm}. We also highlight that every rotated Petz recovery map ${\cal R}_{\cal D(\rho)}^\theta$ is a covariant quantum channel with respect to $L$ and recovers all diagonal elements of the initial state as $({\cal R}^\theta_{{\cal D} (\rho)} \circ {\cal N}_{\rm cov} )  ({\cal D}(\rho)) = {\cal D}(\rho)$. By combining these observations with Theorem~\ref{CovRevThm}, we establish the relationship between the asymmetry loss and the reversibility via a covariant recovery channel. This has been studied in the context of the catalytic transformation of quantum states \cite{Marvian16}.
\begin{corollary} [Recovery of  covariant channels \cite{Marvian16}]
A quantum state is fully recoverable \textit{if and only if} there is no asymmetry loss $\Delta C =0$ through the covariant quantum channel ${\cal N}_{\rm cov}$.
Also, there exists a covariant recovery map that gives the recovery fidelity satisfying 
$$
{\cal F}(\rho, (\bar{{\cal R}}_{{\cal D}(\rho)} \circ {\cal N}_{\rm cov})(\rho)) \geq e^{\Delta C}.
$$
\end{corollary}

We introduce another application of the QFT in the resource theory of asymmetry. We note that although the average amount of asymmetry cannot be increased under a covariant quantum channel, some off-diagonal components $\rho_{ij}$ can be merged \cite{Cwiklinski15, LostaglioX} to get a larger off-diagonal coefficient $|{\cal \rho}_{k' l'}|$ in the output state. We can prove the upper bound for the value of the off-diagonal element using the QFT, which provides more information than the mean value of the asymmetry loss.
\begin{theorem}[Coherence merging bound for a covariant process]
\label{CohMerg} Suppose that the quantum channel ${\cal N}_{\rm cov}$ is covariant with respect to the generator $L$. 
After passing through the channel, the value of the off-diagonal elements is upper bounded by
$$
\left| {\cal N}_{\rm cov}(\rho)_{k'l'} \right| \leq \sum_{\Omega^+_{k'l'}} |\rho_{ij}| e^{-(\delta q_R)_{ij \rightarrow k'l'}}+ \sum_{\Omega^-_{k'l'}} |\rho_{ij}|,
$$
where $\Omega^+_{k'l'}$ and $\Omega^-_{k'l'}$ are the subsets of $\Omega_{k'l'} = \{(i,j) | \lambda_i - \lambda_j = \lambda_{k'} - \lambda_{l'}\}$ with $(\delta q_R)_{ij \rightarrow k'l'} \geq 0$ and $(\delta q_R)_{ij \rightarrow k'l'} <0$ for an arbitrary reference state $\gamma$ commuting with $L$.
\end{theorem}
\noindent The coherence merging inequality in Theorem~\ref{CohMerg} is a generalization of the coherence merging inequalities in quantum thermodynamics \cite{Cwiklinski15, LostaglioX}, as the Theorem can be applied to not only a thermal channel but also any covariant quantum channels.

\begin{figure}[t]
\includegraphics[width=0.9\linewidth]{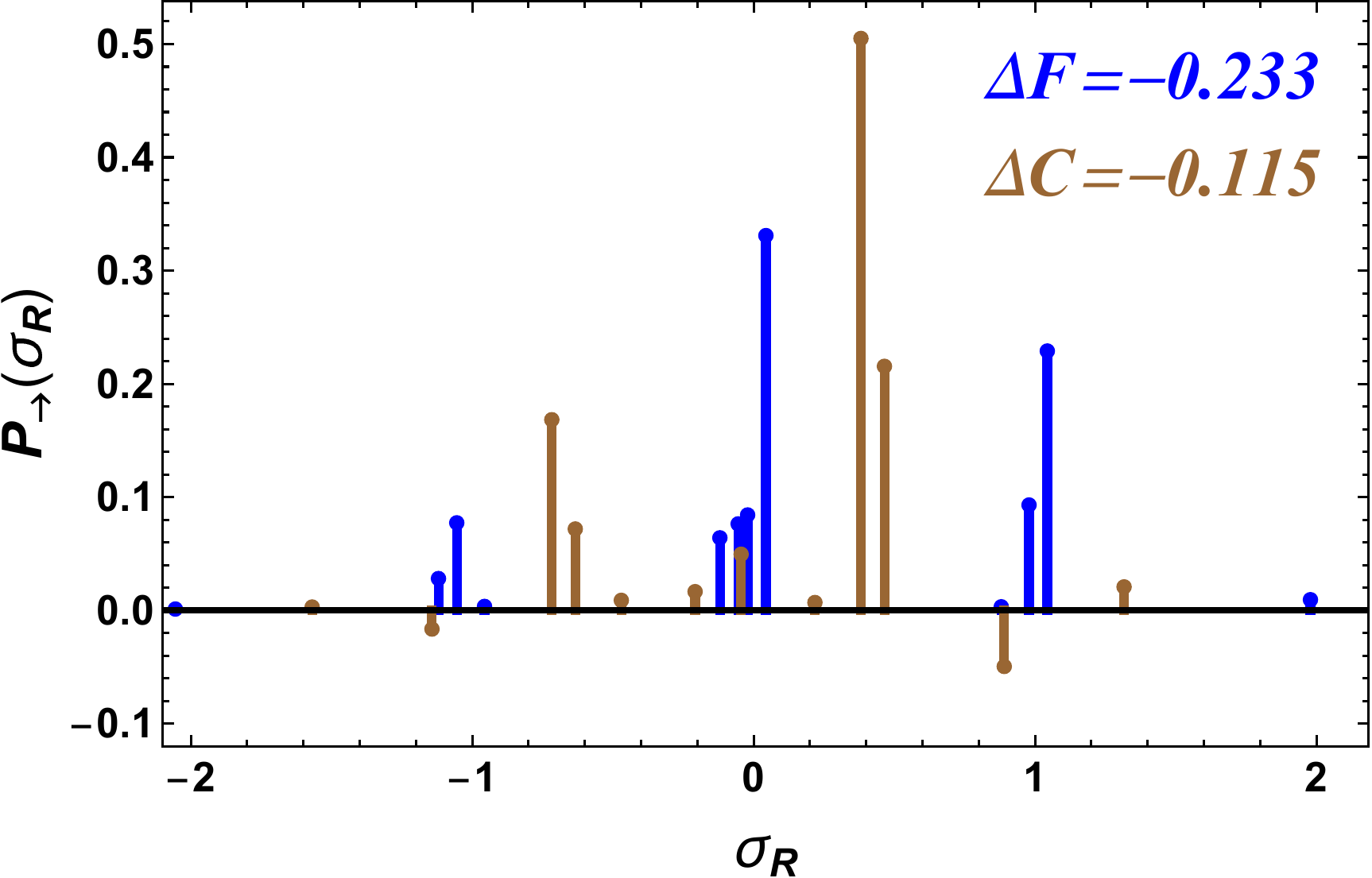}
\caption{ Free energy loss (blue) versus asymmetry loss (brown) in the JC Hamiltonian with thermal noise. The initial state and the parameters are the same as in Fig.~\ref{NegEntPFig}, and $P_\rightarrow(\sigma_R)$ for each case is given by the summation over $\sigma_I$. Both the free energy and coherence loss satisfy the integral QFT: $\langle e^{ \beta \delta F }\rangle = e^{\beta \Delta F_{\rm eq}} =1$ and $\langle e^{ \delta C_R } \rangle = 1$.}
\label{FreeEvsCoh}
\end{figure}
Let us discuss how the different choices of reference states lead to different interpretations of loss albeit through the same quantum channel. For this purpose, we adopt the previously discussed JC Hamiltonian with the parameters $\beta = 1$, $\omega_0 =1$ and $g=0.1$, and the thermal noise term ${\cal L}_{\rm noise}$ with $\Gamma = 0.1$. We set the initial atomic state $\rho = (1/2) \ket{\psi} \bra{\psi} + \mathbb{1}/4$ with $\ket{\psi}= (\ket{g} + \ket{e})/\sqrt{2}$. We note that the dynamics of the system can be interpreted as a thermodynamic channel discussed in the last part of Section~\ref{IntQFTSection}, as well as a covariant channel with respect to the system Hamiltonian $H_a$ studied in Section~\ref{subsec-asymm}. By regarding this channel as a thermodynamic process, we can choose the Gibbs state $\gamma_a$ as the reference state. In this case, the mean entropy production indicates the loss in the free energy $\Delta F = \langle \delta F \rangle =-0.233$, and the reverse channel has the meaning of time-reversal operation in thermodynamics. For the same channel and initial state, another choice of the reference state would be taking the diagonal state ${\cal D}(\rho)$ to focus on the covariant property of the channel. In this case, the corresponding reverse channel would be a covariant recovery channel, and the mean entropy production $\Delta C = \langle \delta C_R \rangle = -0.115$ can be regarded as the dissipated asymmetry in the quantum state through the quantum channel. While the average values of free energy and coherence loss are different, both satisfy the integral QFT, $\langle e^{ \beta \delta F }\rangle = e^{\beta\Delta F_{\rm eq}} =1$ and $\langle e^{ \delta C_R } \rangle = 1$ by taking $\theta = 0$ and noting that $\Delta F_{\rm eq} = 0$ as the Hamiltonian remains the same. Figure~\ref{FreeEvsCoh} shows the distributions $P_\rightarrow (\sigma_R)$ for the free energy and coherence loss. This approach can be utilized to characterize and quantify the loss of resources through a quantum channel in various physical contexts.

\subsection{Fluctuation of entanglement and coherent information}
We apply the QFT to study a stochastic entanglement generation by a local operation and classical communication (LOCC). We consider a LOCC protocol on a pure bipartite state $\ket{\Psi}_{AB}$, which is similar to the construction introduced in Ref.~\cite{Alhambra17}, to study the fluctuation of entanglement. The initial amount of entanglement between $A$ and $B$ is quantified by the entanglement entropy as
$$
E_S(\ket{\Psi}_{AB}) = S(\rho_A) = S(\rho_B),
$$
where $\rho_A$ and $\rho_B$ are the marginal states of the local parties $A$ and $B$, respectively. Now let us perform a local POVM on $B$ with a set of measurement operators $\{K_m\}$, followed by a local unitary operation $V_m$ on $A$ based on the measurement outcome $m$. The resulting state would be another bipartite pure state 
$$| \Phi_m \rangle_{AB} = (V_m \otimes K_m ) \ket{\Psi}_{AB} / \sqrt{P_m},$$
with the probability $P_m = \bra\Psi K_m^\dagger K_m \ket\Psi $ for the outcome $m$. Since the local unitary operation $V_m$ does not change the entanglement entropy of a pure bipartite state, the entanglement difference between the initial and final states is given by $\Delta E_S^m= E_S(|\Phi_m \rangle_{AB}) - E_S(\ket{\Psi}_{AB})$. The monotonicity of entanglement \cite{Vidal00} tells us that the amount of entanglement does not increase under LOCC on average, i.e.,
$$\Delta E_S := \sum_m P_m \Delta E_S^m \leq 0.$$

The QFT in Theorem~\ref{2ndLawThm} shows that the fluctuation of the entanglement loss obeys the balanced equality relation. By introducing the memory state $\ket{m}_M\bra{m}$, the quantum channel ${\cal N}_{\rm LOCC}$ of the LOCC protocol can be described as follows:
$$
\ket{\Psi}_{AB} \bra{\Psi} \xrightarrow{{\cal N}_{\rm LOCC}} \sum_m P_m | \Phi_m \rangle_{AB} \langle \Phi_m | \otimes \ket{m}_M\bra{m},
$$
where $\ket{m}_M$ are orthogonal to each other. We also take the reference state to be ${\mathbb 1}_A \otimes \rho_B$, then its evolution is given by ${\cal N}_{\rm LOCC} ( {\mathbb 1}_A \otimes \rho_B) =  {\mathbb 1}_A \otimes \sum_m P_m \rho_B^m \otimes \ket{m}_M \bra{m}$. Here, $\rho_B^m = {\rm Tr}_A | \Phi_m \rangle_{AB} \langle \Phi_m | $ is the local state of $B$ for the measurement outcome $m$. We note that $\rho_B = \sum_i r_i \ket{i}_A \bra{i}$ and $\rho_B^m = \sum_{k'} r^m_{k'} \ket{m,k'}_B\bra{m,k'}$ by using the Schmidt decompositions $\ket{\Psi}_{AB} = \sum_i \sqrt{r_i} \ket{i}_A \ket{i}_B$ and $\ket{\Phi_m}_{AB} = \sum_{k'} \sqrt{r^m_{k'}} \ket{m, k'}_A \ket{m,k'}_B$. Based on the reference state for the LOCC channel and regarding the initial memory state as $\ket{0}_M\bra{0}$, the single-shot entropy production for the transition $(0,i,j) \rightarrow (m, k', l')$ is given by $\sigma^{0 \rightarrow m}_{0ij \rightarrow m k'l'} = \log \sqrt{r^m_{k'} r^m_{l'}} - \log \sqrt{r_i r_j} + i \left(  \log \sqrt{{r^m_{l'}}/{r^m_{k'}}} - \log \sqrt{{r_i}/{r_j}}\right)$, where $\sqrt{r_i}$ and $\sqrt{r^m_{k'}}$ are the Schmidt coefficients of $\ket{\Psi}_{AB}$ and $\ket{\Phi_m}_{AB}$. The QFT provides the balanced equality condition for the difference between the Schmidt coefficients by defining $(\delta E_S)_{0ij \rightarrow m k'l'} = -\sigma^{0 \rightarrow m}_{0ij \rightarrow m k'l'}$. The integral QFT for the entanglement loss is then obtained from Theorem~\ref{2ndLawThm}
$$
\left\langle e^{ (\delta E_S)_R - i \theta (\delta E_S)_I  } \right\rangle = {\cal F}_\theta,
$$
where ${\cal F}_\theta = {\cal F}(\rho_{AB}, ({\cal R}_{\mathbb{1}_A \otimes \rho_B}^{\theta/2} \circ {\cal N}_{\rm LOCC})(\rho_{AB}))$ is the recovery fidelity of $\rho_{AB} = \ket{\Psi}_{AB} \bra{\Psi}$ by the recovery channel ${\cal R}_{\mathbb{1}_A \otimes \rho_B}^{\theta/2}$. Theorem~\ref{2ndLawThm} leads to the monotonicity of entanglement \cite{Vidal00} $\Delta E_S = -\langle \sigma \rangle \leq 0$.

It is important to note that the Petz recovery map corresponding to the LOCC channel ${\cal N}_{\rm LOCC}$ also belongs to LOCC, and this channel fully recovers the marginal states of the local parties. This observation, combined with Theorem~\ref{CovRevThm}, directly connects the entanglement loss through the LOCC channel ${\cal N}_{\rm LOCC}$ to the recoverability of the entangled state as follows:
\addtocounter{theorem}{-1}
\addtocounter{corollary}{+2}
\begin{corollary}[Recovery of LOCC channels]
\label{Coro-LOCC} A pure bipartite state $\ket{\Psi}_{AB}$ can be fully recovered by the LOCC channel \textit{if and only if} there is no entanglement loss by ${\cal N}_{\rm LOCC}$, i.e., $\Delta E_S = 0$. Also, the entanglement loss is bounded by the average recovery fidelity ${\cal \bar{F}} := \int_{-\infty}^{\infty} d\theta g_0(\theta) {\cal F}_\theta$ as
$$
 \Delta E_S  \leq  \log {\cal \bar{F}}.
$$
\end{corollary}
This result can also be generalized to a bipartite mixed state $\rho_{AB}$. The LOCC protocol ${\cal N}_{\rm LOCC}$ transforms the initial state $\rho_{AB}$ into $\rho^m_{AB} = (V_m \otimes K_m) \rho_{AB} (V_m^\dagger \otimes K_m^\dagger) / P_m$ when the measurement outcome is $m$. We choose the reference state $\gamma ={\mathbb 1}_A \otimes \rho_B$, where $\rho_B$ is the marginal state of $\rho_{AB}$, similar to the case of the pure state. The first-order moment of the entropy production then becomes
$$
\Delta I(A \rangle B) := \sum_m P_m I(A \rangle B)_{\rho^m_{AB}} - I (A \rangle B)_{\rho_{AB}}  =- \langle \sigma \rangle \leq 0,
$$
where $I(A \rangle B)_{\rho_{AB}} = S(\rho_B) - S(\rho_{AB})$ is the coherent information \cite{Schumacher96, Horodecki05}. 
When $\rho_{AB}$ is pure, the coherent information $I(A \rangle B)$ is reduced to the entanglement entropy $E_S$. As the coherent information quantifies the amount of quantum correlation between the two parties, the QFT in Eq.~(\ref{IntQFT2}) captures the fluctuation of correlation loss through the LOCC protocol. In particular, this result can be utilized to investigate the fluctuation relation of quantum information quantities under feedback control \cite{Sagawa10, Naghiloo18}. It would also be interesting to explore the case when the system and bath are initially correlated \cite{Romero04}. Although the dynamics of the system may not be a linear channel \cite{Romero04}, the dynamics of the entire system including the bath can be described by a CPTP map so that the information exchange between the system and bath can be studied in our QFT framework. In this way, thermodynamic quantities such as quantum heat and work can be coherently combined to both quantum and classical information quantities involved in the Maxwell demon or Laundauer's erasure in a unified framework.

\section{Remarks} 
\label{Sec-Conclusion}
We have established a general framework of QFT by showing that it is always possible for any linear quantum channel to find the symmetry between the forward and backward probabilities. In our formulation, the Petz recovery map can be understood as a family of reverse quantum channels, and entropy production in conventional FTs can be generalized into the quantum regime. The effect of coherences has been taken into account in two different aspects: coherences in the quantum system and coherent transitions by the channel. Coherences in the quantum system lead to the modification of the FT as quantum corrections are required in the conventional fluctuation relations. We have seen that the significant differences arise when the quantum channel induces off-diagonal transitions. By introducing a complex-valued quantum entropy production, the transitions between both diagonal and off-diagonal elements of the system density matrix can be understood by a single formula. The imaginary part of the quantum entropy production emerges at the point where nontrivial coherence transfer occurs, which may imply the broken symmetry. We also provide concrete examples of a two-level atom to explore the coherence transfers induced by the dynamics and the emergence of imaginary entropy productions. We highlight that our approach can be applied to any physical systems and dynamics described by a CPTP channel, ranging from a single unitary operation to complicated Lindblad equations.

Another important progress in this work is finding a direct connection between the QFT and the quantum data processing inequality. While both real and imaginary entropy productions are essential to derive the second law for a quantum channel from the QFT, only the real entropy production contributes to the mean entropy production represented by the quantum relative entropy difference. Our results provide a toolkit to analyze the dissipation of  quantum resources through a quantum channel. As the reference state to the Petz recovery map can be chosen in various ways, our QFT allows investigating a given quantum channel from various angles, e.g. energy, coherence, entanglement, by providing a refined statistics of the dissipated resources. The relationship between the QFT of dissipated quantum resources and the monotonicity of the resource measure can be compared to that between the classical FTs and the second law of thermodynamics.

Yet there are possible applications and open-problems which can be studied in future research. In quantum thermodynamics, our approach can be useful to generalize the FTs with feedback control \cite{Sagawa10} into the quantum regime by fully understanding the fluctuation relation including quantum information exchange between the system and quantum memory \cite{Sagawa12}. Finding the deeper physical meaning of the imaginary information exchange or entropy production would be another interesting question, and it may be useful to distinguish a quantum channel by coherent and incoherent components. Quantum error correction protocols could benefit from this; the loss caused by a noisy quantum channel can be analyzed with greater details, and the resource-efficient quantum error correction protocols can be developed in a covariant way \cite{Faist19, Woods19}. Another interesting direction of future research would be developing a generalized measurement protocol \cite{Dorner13, Roncaglia14} to directly measure both real and imaginary entropy productions in a noninvasive way.

\acknowledgements
The authors thank David Jennings, Mario Berta, and Erick Hinds Mingo for helpful discussions. This work is supported by the KIST Open Research Program, the QuantERA ERA-NET within the EU's Horizon 2020 Programme, and the EPSRC  (EP/R044082/1). M.S.K. acknowledges the Royal Society and Samsung GRO grant for support.

\appendix
\section{Proof of Theorem~\ref{PureFT}:}
Recall that the reverse channel is expressed as ${\cal R}_\gamma  = \JJ{\gamma}{\frac{1}{2}} \circ {\cal N}^\dagger \circ \JJ{{\cal N}(\gamma)}{-\frac{1}{2}}$. Then we have
$$
\begin{aligned}
&T(\ket{\psi} \rightarrow \ket{\phi'})\\ &= \bra{\phi'} {\cal N} (\ket{\psi} \bra{\psi}) \ket{\phi'} \\
&= \langle \phi' | {\cal N}(\gamma)^{\frac{1}{2}} (\JJ{{\cal N}(\gamma)}{-\frac{1}{2}} \circ {\cal N} \circ \JJ{\gamma}{\frac{1}{2}}) ( \gamma^{-\frac{1}{2}} |\psi \rangle \langle \psi | \gamma^{-\frac{1}{2}}  ) {\cal N}(\gamma)^{\frac{1}{2}} |\phi'\rangle\\
&= \langle \tilde\phi' | (\JJ{{\cal N}(\gamma)}{-\frac{1}{2}} \circ {\cal N} \circ \JJ{\gamma}{\frac{1}{2}}) (| \tilde{\psi} \rangle \langle \tilde{\psi} |) | \tilde \phi'\rangle \bra{\psi} \gamma^{-1} \ket{\psi} {\bra{\phi'} {\cal N}(\gamma) \ket{\phi'}} \\
&= \langle \tilde{\psi} | (\JJ{\gamma}{\frac{1}{2}} \circ {\cal N}^\dagger \circ \JJ{{\cal N}(\gamma)}{-\frac{1}{2}}) (| \tilde \phi' \rangle \langle \tilde \phi' |) | \tilde{\psi}\rangle \bra{\psi} \gamma^{-1} \ket{\psi} {\bra{\phi'} {\cal N}(\gamma) \ket{\phi'}} \\
&= \tilde T ( | \tilde{\psi} \rangle \leftarrow | \tilde \phi' \rangle) \bra{\psi} \gamma^{-1} \ket{\psi} {\bra{\phi'} {\cal N}(\gamma) \ket{\phi'}},
\end{aligned}
$$
which completes the proof. \hfill $\qed$

\section{Recovery map for a Lindblad equation}
We show that the reverse process of the forward Lindblad equation 
$$
{\cal L} (\rho) = - \frac{i}{\hbar} [H_t, \rho] + \sum_n \left( L_n \rho L_n^\dagger - \frac{1}{2} \{ L_n^\dagger L_n, \rho\} \right)
$$
is given by the following backward Lindblad equation
$$
{\cal \tilde{L}} (\rho) = - \frac{i}{\hbar} [\tilde H_t, \rho] + \sum_n \left( \tilde{L}_n \rho \tilde{L}_n^\dagger - \frac{1}{2} \{ \tilde{L}_n^\dagger \tilde{L}_n, \rho \}\right)
$$
by applying the Petz recovery map for infinitesimal time interval $dt$. For mathematical simplicity, we set $\hbar = 1$. Note that the reverse channel for infinitesimal time interval $dt$ is given by 
$$
{\cal R}_{t \leftarrow t+dt} = \JJ{\gamma_{t}}{{1}/{2}} \circ (\mathbb{1} + {\cal L}^\dagger dt) \circ \JJ{\gamma_{t+dt}}{-{1}/{2}},
$$
where ${\cal L}^\dagger (A) = i [ H_t, A] + \sum_n\big(  L_n^\dagger A L_n - (1/2) \{ L^\dagger_n L_n, A \}  \big)$. By taking the first order of $dt$, we get
$$
\begin{aligned}
{\cal R}_{t \leftarrow t+dt} (\rho) = \rho &- (\dot{G}G^{-1} \rho + \rho G^{-1} \dot{G}) dt\\
& + G {\cal L}^\dagger (G^{-1} \rho G^{-1}) G dt,
 \end{aligned}
$$
where we denote $G = \gamma_t^{1/2}$ and its time derivative $\dot{G}$  for mathematical simplicity. We note that $G G^{-1} = \mathbb 1$, which leads to $\dot{(G^{-1})} = -G^{-1} \dot{G}G^{-1}$.
The last term of ${\cal R}_{t \leftarrow t+dt}$ can be expanded as
$$
\begin{aligned}
G {\cal L}^\dagger (G^{-1} \rho G^{-1}) G &= i (GH_t G^{-1} \rho - \rho G^{-1} H_t G) \\
&\quad+\sum_n \bigg( G L_n^\dagger G^{-1} \rho G^{-1} L_n G \\
&\quad - \frac{1}{2} G L_n^\dagger L_n G^{-1} \rho + \rho G^{-1} L_n^\dagger L_n G \bigg).
\end{aligned}
$$
By defining $\tilde L_n = G L_n^\dagger G^{-1}$, we can rewrite the above equation as 
$$
\begin{aligned}
G {\cal L}^\dagger (G^{-1} \rho G^{-1}) G &= \sum_n \left( \tilde L_n \rho \tilde L_n^\dagger - \frac{1}{2} \{ \tilde L_n^\dagger \tilde L_n , \rho \}\right)\\
&\quad + i (GH_t G^{-1} \rho - \rho G^{-1} H_t G) \\
&\quad + \frac{1}{2} \sum_n \{ \tilde L_n^\dagger \tilde L_n, \rho \}\\
&\quad+\sum_n \bigg ( G L_n^\dagger G^{-1} \rho G^{-1} L_n G \\
&\quad - \frac{1}{2} G L_n^\dagger L_n G^{-1} \rho + \rho G^{-1} L_n^\dagger L_n G \bigg).
\end{aligned}
$$
Meanwhile, we also note that $G = \gamma_t^{1/2}$ then
$$
\sum_n L_n G G L_n^\dagger = {\cal L}(\gamma_t) + i [H_t, \gamma_t] + \frac{1}{2} \sum_n \{ L_n^\dagger L_n, \gamma \},
$$
where ${\cal L}(\gamma_t) = \dot{ \gamma_t } = G \dot G + \dot G G$.
This leads to
$$
\begin{aligned}
\sum_n \tilde L_n^\dagger \tilde L_n &= G^{-1} \bigg( G\dot G + \dot G G + i [H_t, GG] \\
&\qquad \qquad + \frac{1}{2} \sum_n \{ L_n^\dagger L_n, GG \}  \bigg) G^{-1}\\
&= \dot G G^{-1} + G^{-1} \dot G + i G^{-1} H_t G - i G H_t G^{-1} \\
&\quad+ \frac{1}{2} \sum_n \left( G^{-1} L_n^\dagger L_n G + G L_n^\dagger L_n G^{-1} \right).
\end{aligned}
$$
By combining these altogether, we finally get 
$$
\begin{aligned}
{\cal R}_{t \leftarrow t+dt} (\rho) &= \frac{i}{2} \bigg[ G H_t G^{-1} + i  \dot G G^{-1} +\frac{i}{2} \sum_n G L_n^\dagger L_n G^{-1} , \rho \bigg] \\
&~+\frac{i}{2} \bigg [ G^{-1} H_t G - i  G^{-1} \dot G -\frac{i}{2} \sum_n G^{-1} L_n^\dagger L_n G , \rho \bigg] \\
&~+\sum_n \left( \tilde L_n \rho \tilde L_n^\dagger - \frac{1}{2} \{ \tilde L_n^\dagger \tilde L_n , \rho \}\right)\\
&=- i [\tilde H_t, \rho] + \sum_n \left( \tilde{L}_n \rho \tilde{L}_n^\dagger - \frac{1}{2} \{ \tilde{L}_n^\dagger \tilde{L}_n, \rho \}\right),
\end{aligned}
$$
where the Hamiltonian for the reverse process $\tilde H_t$ is defined as
$$
\tilde H_t = -\frac{1}{2} \big(G H_t G^{-1} + i  \dot G G^{-1} +\frac{i}{2} \sum_n G L_n^\dagger L_n G^{-1}\big) + {\rm h.c.}.
$$
After recovering $\hbar$, we obtain the expression in Eq.~(\ref{MarkovRev}). This result can also be generalized for the rotated Petz recovery map by taking $G = \gamma_t^{1/2+i\theta}$ and taking into account for $G^\dagger = \gamma_t^{1/2-i\theta}$ as its Hermitian conjugate.

\section{Proof of Theorems~\ref{CFT} and \ref{RotatedCFT}:}
We prove Theorem~\ref{RotatedCFT} as follows:
$$
\begin{aligned}
&P_\rightarrow (\sigma) e^{-\sigma_R + 2 i \theta \sigma_I } \\
&= \sum_{\mu, i,j} \sum_{\nu',k',l'} P^{\mu, {\nu'}}_{ij,k'l'} \delta \left(\sigma - \sigma^{\mu \rightarrow {\nu'}}_{ij \rightarrow k'l'}\right) e^{-\sigma_R + 2 i \theta \sigma_I }\\
&= \sum_{\mu, i,j} \sum_{\nu',k',l'} p_\mu T_{ij \rightarrow k'l'} e^{-\sigma_R + 2 i \theta \sigma_I } \delta \left(\sigma - \sigma^{\mu \rightarrow {\nu'}}_{ij \rightarrow k'l'}\right)\\
&= \sum_{\mu, i,j} \sum_{\nu',k',l'} p'_{\nu'} \tilde T^*_{ij \leftarrow k'l'} e^{i \theta \log \left( \frac{r_j r'_{k'}}{r_i r'_{l'}} \right)} \delta \left(\sigma - \sigma^{\mu \rightarrow {\nu'}}_{ij \rightarrow k'l'}\right)\\
&= \sum_{\mu, i,j} \sum_{\nu',k',l'} p'_{\nu'} \tilde T_{ij \leftarrow k'l'} e^{i \theta \log \left( \frac{r_i r'_{l'}}{r_j r'_{k'}} \right)} \delta \left(\sigma - \sigma^{\mu \rightarrow {\nu'}}_{ji \rightarrow l'k'}\right)\\
&= \sum_{\mu, i,j} \sum_{\nu',k',l'} p'_{\nu'} \tilde T_{ij \leftarrow k'l'} e^{i \theta \log \left( \frac{r_i r'_{l'}}{r_j r'_{k'}} \right)} \delta \left(\sigma^* + \sigma^{\mu \leftarrow {\nu'}}_{ij \leftarrow k'l'}\right)\\
&= P^\theta_\leftarrow (-\sigma^*),
\end{aligned}
$$
by using the fact that $\tilde T^*_{ij \leftarrow k'l'} = T_{ji \leftarrow l'k'}$ and $\sigma^{\mu \rightarrow {\nu'}}_{ji \rightarrow l'k'} = - [\sigma^{\mu \leftarrow {\nu'}}_{ij \leftarrow k'l'}]^*$. Note that Theorem~\ref{CFT} is the special case of Theorem~\ref{RotatedCFT} with $\theta=0$ as $P_\rightarrow (\sigma_R + i \sigma_I) e^{-\sigma_R} = P_\leftarrow (-\sigma_R + i \sigma_I)$. By summing over all $\sigma_I$, we achieve Theorem~\ref{CFT}.
\hfill $\qed$

\section{Obtaining the quasi-probability distribution from a two-point POVM}
\label{appx:TPPOVM}
We demonstrate that the TPM quasi-probability $P^{\mu, \nu'}_{ij,k'l'}$ can be obtained from the distribution of a two-point POVM 
$$
P_\rightarrow (m,m') = \Tr[ M'_{m'} {\cal N} (M_m \rho M_m^\dagger) {M'}^\dagger_{m'}],
$$
where $\{ M_m \} = \{ \frac{1}{\sqrt{d}}\Pi_i \Pi_{\psi_\mu} , \frac{1}{\sqrt{2d}}(\Pi_i +\Pi_j)\Pi_{\psi_\mu}, \frac{1}{\sqrt{2d}}(\Pi_i + i\Pi_j)\Pi_{\psi_\mu} \}$ and $\{ M'_{m'} \} = \{ \frac{1}{\sqrt{d}} \Pi_{\phi_{\nu'}} \Pi_{k'} , \frac{1}{\sqrt{2d}} \Pi_{\phi_{\nu'}} (\Pi_{k'} +\Pi_{l'}), \frac{1}{\sqrt{2d}} \Pi_{\phi_{\nu'}} (\Pi_{k'} + i \Pi_{l'}) \}$ for every possible $\mu, i, j$ and $\nu', k', l'$ satisfying $i<j$ and $k' <l'$. To do this, we show that $P^{\mu, \nu'}_{ij,k'l'}$ can be expressed in terms of $P_\rightarrow (m,m')$ for given pairs of $(\mu, i,j)$ and $(\nu', k',l')$. Without loss of generality, we assume that $i,j \in \{ 0, 1\}$ and $k', l' \in \{0', 1'\}$. For mathematical simplicity, we define
$$
P(a,b') := \Tr[ M'_{(\nu', b')} {\cal N} (M_{(\mu, a)} \rho M_{(\mu, a)}^\dagger) {M'}^\dagger_{(\nu',b')}]
$$
for fixed values of $\mu$ and $\nu'$, where
$$
\begin{aligned}
M_{(\mu, 0)} &:= \frac{1}{\sqrt{d}}\Pi_0 \Pi_{\psi_\mu} \\
M_{(\mu, 1)} &:= \frac{1}{\sqrt{d}}\Pi_1 \Pi_{\psi_\mu} \\
M_{(\mu, +)} &:= \frac{1}{\sqrt{2d}} (\Pi_0 + \Pi_1) \Pi_{\psi_\mu} \\
M_{(\mu, \times)} &:= \frac{1}{\sqrt{2d}} (\Pi_0 + i \Pi_1) \Pi_{\psi_\mu}
\end{aligned}
$$
and 
$$
\begin{aligned}
M'_{(\nu', 0')} &:= \frac{1}{\sqrt{d}}  \Pi_{\phi_{\nu'}} \Pi_{0'} \\
M'_{(\nu', 1')} &:= \frac{1}{\sqrt{d}}  \Pi_{\phi_{\nu'}} \Pi_{1'}\\
M'_{(\nu', +')} &:= \frac{1}{\sqrt{2d}}  \Pi_{\phi_{\nu'}} (\Pi_{0'} + \Pi_{1'}) \\
M'_{(\nu', \times')} &:= \frac{1}{\sqrt{2d}}  \Pi_{\phi_{\nu'}} (\Pi_{0'} + i \Pi_{1'}).
\end{aligned}
$$
First, we note that 
$$
P^{\mu, \nu'}_{ii,k'k'} = d^2 P(i ,k')
$$
with $i = 0,1$ and $k' = 0', 1'$.
For $i = j$ and $k' \neq l'$, we define
$$
Q( a, b') := P(a,b') - \frac{1}{2} \sum_{k'=0',1'} P (a,k')
$$
for $a \in \{0,1\}$ and $b' \in \{ +' , \times' \}$.
We then obtain
$$
\begin{aligned}
P^{\mu, \nu'}_{ii,0'1'} &= d^2 [ Q(i, +') + i Q(i, \times') ] \\
P^{\mu, \nu'}_{ii,1'0'} &= d^2 [ Q(i, +') - i Q(i, \times') ].
\end{aligned}
$$
Similarly, for $i \neq j$ and $k' = l'$, we obtain
$$
\begin{aligned}
P^{\mu, \nu'}_{01,k'k'} &= d^2 [ Q(+,k') + i Q(\times, k') ] \\
P^{\mu, \nu'}_{10,k'k'} &= d^2 [ Q(+,k') - i Q(\times, k') ],
\end{aligned}
$$
where
$$
Q( a, b') := P(a,b') - \frac{1}{2} \sum_{i=0,1} P (i,b')
$$
for $a \in \{ +, \times \}$ and $b' \in \{ 0' , 1'\}$.
In order to obtain the TPM quasi-probability for $i \neq j$ and $k' \neq l'$, we additionally define
$$
\begin{aligned}
Q( a, b') &:= P(a,b') \\
& \quad - \frac{1}{2} \left[ \sum_{i=0,1} Q(i,b') - \sum_{k'=0',1'}Q (a,k') \right] - \frac{1}{4} \bar{P}
\end{aligned}
$$
for $a \in \{ +, \times \}$ and $b \in \{ +', \times'\}$, where $\bar{P} = \sum_{i=0,1} \sum_{k'=0',1'} P(i,k')$.
Finally, we obtain
$$
\left(
\begin{matrix}
P^{\mu, \nu'}_{010'1'} \\ P^{\mu, \nu'}_{011'0'} \\ P^{\mu, \nu'}_{100'1'} \\ P^{\mu, \nu'}_{101'0'}
\end{matrix}
\right)
= d^2
\left(
\begin{matrix}
1 & i & i & -1 \\ 1 & -i & i & 1 \\ 1 & i & -i & 1 \\ 1 & -i & -i & -1
\end{matrix}
\right)
\left(
\begin{matrix}
Q(+,+') \\ Q(+,\times')\\ Q(\times,+') \\ Q(\times,\times')
\end{matrix}
\right),
$$
thus we conclude that every element of $P^{\mu, \nu'}_{ij, k'l'}$ is expressed in terms of $P(m, m')$. Then, $P_\rightarrow(\sigma)$ for both real and imaginary $\sigma$ can be obtained from the TPM quasi-probability distribution $P^{\mu, \nu'}_{ij,k'l'}$.

\section{Proof of Theorem~\ref{CovThm}:}
We first show that if a quantum channel ${\cal N}$ is covariant with respect to the reference state, i.e.
$$\JJ{{\cal N}(\gamma)}{-i \theta} \circ {\cal N} \circ \JJ{\gamma}{i \theta}= {\cal N},$$
all of its rotated Petz recovery maps ${\cal R}_\gamma^\theta$ are the same with ${\cal R}_\gamma$. Note that $\JJ{\gamma}{i \theta}$ and $\JJ{ {\cal N}(\gamma)}{-i \theta}$ are unitary processes; therefore, the adjoint map ${\cal N}^\dagger$ satisfies
$$
{\cal N}^\dagger = \left( \JJ{{\cal N}(\gamma)}{-i\theta} \circ {\cal N} \circ \JJ{\gamma}{i\theta} \right)^\dagger = {\JJ{\gamma}{-i\theta} \circ {\cal N}^\dagger \circ \JJ{{\cal N}(\gamma)}{i\theta}.}
$$
Therefore, the rotated Petz recovery map is given by
$$
\begin{aligned}
{\cal R}_\gamma^\theta (\rho) &= ( \JJ{\gamma}{\frac{1}{2}+i\theta} \circ {\cal N}^\dagger \circ \JJ{{\cal N}(\gamma)}{-\frac{1}{2} - i\theta}  ) (\rho)\\
&= ( \JJ{\gamma}{\frac{1}{2}} \circ \JJ{\gamma}{i \theta} \circ {\cal N}^\dagger \circ \JJ{{\cal N}(\gamma)}{- i \theta} \circ \JJ{{\cal N}(\gamma)}{-\frac{1}{2}} ) (\rho)\\
&= ( \JJ{\gamma}{\frac{1}{2}}  \circ {\cal N}^\dagger \circ \JJ{{\cal N}(\gamma)}{-\frac{1}{2}} ) (\rho)\\
&={\cal R}_\gamma (\rho).
\end{aligned}
$$

Now we prove the remaining part of Theorem~\ref{CovThm}. In terms of the elements in the transition matrix, every covariant quantum process obeys
$$
T_{ij \rightarrow k'l'} e^{ i \theta \log \sqrt{ \frac{r_j r'_{k'}}{r_i r'_{l'}}} } =T_{ij \rightarrow k'l'}.
$$
Multiplying both sides by $e^{-i \theta \sigma_I}$ followed by the integration over $\theta$ leads to 
\begin{equation}
T_{ij \rightarrow k'l'}  \delta \left(\sigma_I - \log \sqrt{\frac{r_i r'_{l'}}{ r_j r'_{k'}}} \right) = T_{ij \rightarrow k'l'} \delta(\sigma_I).
\end{equation}
Therefore, the probability distribution of entropy production is given by 
$$
P_\rightarrow(\sigma_R + i \sigma_I) = P_\rightarrow(\sigma_R) \delta (\sigma_I)
$$
and the QFT with real values of $\sigma$ is given by 
$$
\frac{P_\rightarrow(\sigma)}{ P_\leftarrow(-\sigma)} = e^{\sigma},
$$
where the parameter $\theta$ is omitted since every rotated recovery map is identical to the Petz recovery map ${\cal R}_\gamma$.\hfill $\qed$

\section{Imaginary entropy production and symmetry breaking}
We note that the non-vanishing point of the entropy production distribution can be written as
$$
\begin{aligned}
&P_\rightarrow(\sigma_R + i \sigma_I) \\
&=\sum_{\mu,\nu'} \sum_{\Delta q_R}  \delta (\sigma_R - \delta s^{\mu \rightarrow \nu'} + \Delta q_R) \\  
&~~\times \sum_{i,j,k',l'}  P^{\mu, \nu'}_{ij,kl} \delta(\Delta q_R - (\delta q_R)_{ij \rightarrow k'l'} )\delta(\sigma_I + (\delta q_I)_{ij \rightarrow k'l'} )\\
&=\sum_{\mu,\nu'} \sum_{\Delta q_R}  p_\mu  \delta (\sigma_R - \delta s^{\mu \rightarrow \nu'} + \Delta q_R) \\  
&\qquad \qquad \times \frac{1}{(2\pi)^2}\int d^2 \xi e^{i\xi_R \Delta q_R} e^{-i\xi_I \sigma_I} \chi_{\mu \nu'} (\xi_R, \xi_I),
\end{aligned}
$$
where
$$
\begin{aligned}
&\chi_{\mu\nu'} (\xi_R, \xi_I) \\
&\quad :=\langle{\phi'_{\nu'}} |{\cal N}(\gamma)^{\frac{i(\xi_R - \xi_I)}{2}}{\cal N} ({\gamma^{\frac{-i(\xi_R - \xi_I)}{2}} }\ket{\psi_\mu} \bra{\psi_\mu} {\gamma^{\frac{-i(\xi_R + \xi_I)}{2}} })  \\
&\qquad \qquad {\cal N} ( \gamma)^{\frac{i(\xi_R + \xi_I)}{2}} \ket{\phi'_{\nu'}}.
\end{aligned}
$$
Then the contribution from the transition between the rotated eigenstates  comes from $\xi_R = 0$ and $\xi_I = \theta$ as $\int_{-\infty}^\infty  d\theta e^{-i\theta \sigma_I} \chi_{\mu \nu'} (0, \theta)$ is the Fourier transform of $T(U(\theta/2) \ket{\psi_\mu} \rightarrow V(\theta/2) \ket{\phi'_{\nu'}})$.

\section{Proof of Theorem~\ref{2ndLawThm}:} We first show the equivalent expression of the integral QFT in terms of the rescaling maps:
$$
\begin{aligned}
&\langle e^{-\sigma_R + i \sigma_I \theta} \rangle\\
 &= \sum_{\mu, i,j} \sum_{\nu',k',l'}  P^{\mu, {\nu'}}_{ij,k'l'} e^{-(\sigma_R)^{\mu \rightarrow {\nu'}}_{ij \rightarrow k'l'} + i \theta (\sigma_I)^{\mu \rightarrow {\nu'}}_{ij \rightarrow k'l'} } \\
&= \sum_{\mu, i,j} \sum_{\nu',k',l'}  p_\mu T_{ij \rightarrow k'l'}\left( \frac{p'_{\nu'}}{p_\mu} \right) \sqrt{\frac{r_i r_j}{r'_{k'} r'_{l'}}} e^{i \theta \log \sqrt{ \frac{r_j r'_{k'}}{r_i r'_{l'}}}} \\
&= \Tr \big[ {\cal N}(\rho) {\cal N}(\gamma)^{-\frac{1-i\theta}{2}} {\cal N} ( \gamma^{\frac{1-i\theta}{2}} \rho^{-\frac{1}{2}}  \rho \rho^{-\frac{1}{2}} \gamma^{\frac{1+i\theta}{2}} )  {\cal N}(\gamma)^{-\frac{1+i\theta}{2}}  \big]\\
&=\Tr \left[ \left( {\cal J}_{{\cal N}(\rho)}^{\frac{1-i\theta}{2}} \circ {\cal J}_{{\cal N}(\gamma)}^{-\frac{1-i\theta}{2}}\circ {\cal N} \circ {\cal J}_\gamma^{\frac{1-i\theta}{2}}  \circ {\cal J}_\rho^{-\frac{1-i\theta}{2}}\right) (\rho) \right]
\end{aligned}
$$
for every $\theta$.
Then we can rewrite the last expression as
$$
\begin{aligned}
&\Tr \left[ \left( {\cal J}_{{\cal N}(\rho)}^{\frac{1-i\theta}{2}} \circ {\cal J}_{{\cal N}(\gamma)}^{-\frac{1-i\theta}{2}}\circ {\cal N} \circ {\cal J}_\gamma^{\frac{1-i\theta}{2}}  \circ {\cal J}_\rho^{-\frac{1-i\theta}{2}}\right) (\rho) \right]\\
&=\Tr[ \JJ{\rho}{-1/2}(\rho) (\JJ{\gamma}{\frac{1+i\theta}{2}} \circ {\cal N}^\dagger \circ \JJ{{\cal N}(\gamma)}{-\frac{1+\theta}{2}} ) ( {\cal N}(\rho))  ]\\
&=\Tr[ \Pi_{\rho} ({\cal R}_\gamma^{\theta/2} ({\cal N}(\rho)))]\\
&=\kappa_\theta,
\end{aligned}
$$
where $\Pi_{\rho} ( \cdot )$ is the projection operator onto the support of $\rho$. This is due to the fact that $\JJ{\rho}{-\frac{1}{2}}(\rho)$ can be defined only in the Hilbert space spanned by the eigenvectors of $\rho$ having nonzero eigenvalues.

Meanwhile, the expectation values of the first-order moment are
$$
\begin{aligned}
\langle \sigma_R \rangle &= \sum_{\mu, i,j} \sum_{\nu',k',l'}  P^{\mu, {\nu'}}_{ij,k'l'} (\sigma_R)^{\mu \rightarrow {\nu'}}_{ij \rightarrow k'l'} \\
&= \sum_{\mu, i,j} \sum_{\nu',k',l'}  P^{\mu, {\nu'}}_{ij,k'l'}  \left[ \log \left( \frac{p_\mu}{p'_{\nu'}} \right) + \log \sqrt{\frac{r'_{k'} r'_{l'}}{r_i r_j}} \right] \\
&= \sum_\mu p_\mu \log p_\mu - \sum_i \Pi_i \rho \Pi_i \log r_i \\
&\qquad- \sum_{\nu'} p'_{\nu'} \log p'_{\nu'} + \sum_{k'} \Pi_{k'} {\cal N}(\rho) \Pi_{k'} \log r'_{k'} \\
&=S(\rho || \gamma) - S({\cal N}(\rho) ||{\cal N}(\gamma))
\end{aligned}
$$
and
$$
\begin{aligned}
\langle \sigma_I \rangle &= \sum_{\mu, i,j} \sum_{\nu',k',l'}  P^{\mu, {\nu'}}_{ij,k'l'} (\delta q_I)_{ij \rightarrow k'l'}\\
&= \sum_{\mu, i,j} \sum_{\nu',k',l'} P^{\mu, {\nu'}}_{ij,k'l'} \log \sqrt{\frac{r_i r'_{l'}}{r_j r'_{k'}}}\\
&= \frac{1}{2} \bigg[ \sum_i \Pi_i \rho \Pi_i \log r_i - \sum_j \Pi_j \rho \Pi_j \log r_j \\
& \qquad \sum_{k'} \Pi_k' {\cal N}(\rho) \Pi_{k'} \log r'_{k'} - \sum_{l'} \Pi_{l'} {\cal N}(\rho) \Pi_{l'} \log r'_{l'} \bigg]\\
&=0,
\end{aligned}
$$
by using the marginal distribution of the TPM quasi-probability distribution as shown in Eq.~(\ref{Marginal}).
Therefore, the first-order moment is given by the difference between the quantum relative entropy $\langle \sigma \rangle = \langle \sigma_R \rangle - i\theta \langle \sigma_I \rangle = S(\rho || \gamma) - S({\cal N}(\rho) ||{\cal N}(\gamma))$ for every $\theta$.

In order to obtain the generalized second law, we utilize the following multivariate trace inequality recently proven by Sutter, Berta, and Tomamichel \cite{Sutter}, which generalizes the Golden-Thompson and  Araki-Lieb-Thirring inequalities for multiple Hermitian matrices $H_k$:
\begin{equation}
\label{BertaIneq}
\log \norm{ \exp\left({\sum_{k=1}^N H_k }\right) }{p} \leq \int_{-\infty}^\infty d\theta g_0 (\theta) \log \norm{ \prod_{k=1}^N  e^{(1- i\theta) H_k}}{p}.
\end{equation}
Here, $\Vert A \Vert_p := \left[ \Tr \left( \sqrt{A^\dagger A} \right)^p \right]^{\frac{1}{p}}$ is the Schatten $p$-norm with $p\geq1$, and $g_0(\theta) = \frac{\pi/2}{\cosh(\pi \theta) + 1}$. We choose $p=2$ with $H_1 = \frac{1}{2} U^\dagger [ \log {\cal N}(\rho) \otimes {\mathbb 1}_E] U$, $H_2 = -\frac{1}{2} U^\dagger [ \log {\cal N}(\gamma) \otimes {\mathbb 1}_E] U$, $H_3 = \frac{1}{2} \log \gamma$, $H_4 = -\frac{1}{2} \log \rho$, and $H_5 = \frac{1}{2} \log \rho$, where $U$ is the isometry which leads to the Stinespring dilation of the quantum channel ${\cal N}$ as  $\Tr_E U \rho U^\dagger = {\cal N} (\rho)$.
The right-hand-side becomes negative by observing that $\norm{ \prod_{k=1}^5  e^{(1- i\theta) H_k}}{2}^2 \leq 1$ from the integral QFT $\langle e^{-\sigma_R + i \sigma_I \theta} \rangle = \kappa_\theta \leq 1$.
Therefore, we have 
$$
{\rm (RHS)} = \int_{-\infty}^\infty d\theta g_0 (\theta) \log \kappa_\theta \leq 0.
$$ba
Meanwhile, the left-hand-side can be calculated as
$$
\begin{aligned}
&\log \norm{ \exp\left({\sum_{k=1}^5 H_k }\right) }{2}\\
& = \frac{1}{2} \log \Tr \big[ \exp\{ U^\dagger (\log {\cal N}(\rho) \otimes {\mathbb 1}_E) U \\ & \qquad \qquad - U^\dagger (\log {\cal N}(\gamma) \otimes {\mathbb 1}_E) U 
  + \log \gamma  - \log \rho + \log \rho \} \big].
\end{aligned}
$$
The Peierls-Bogoliubov inequality 
$$
\Tr \left[ e^F e^G \right] \geq \Tr \left[ e^{ F + G} \right] \geq \exp \left\{\Tr \left[ F e^G \right] \right\},
$$
which holds for Hermitian matrices $F$ and $G$ and $\Tr [ e^G ] = 1$ leads to
$$
\begin{aligned}
({\rm LHS})
&\geq \frac{1}{2} \Tr \big[ \rho \{ U^\dagger (\log {\cal N}(\rho) \otimes {\mathbb 1}_E) U \\ & \qquad\qquad - U^\dagger (\log {\cal N}(\gamma) \otimes {\mathbb 1}_E) U 
+ \log \gamma  - \log \rho \} \big]\\
&\geq \frac{1}{2} \Tr \big[ {\cal N}(\rho) \{ \log {\cal N}(\rho) - \log {\cal N}(\gamma) \} \\ &\qquad\qquad+ \rho \{ \log \gamma  - \log \rho \} \big]\\
& = \frac{1}{2} \left[ S({\cal N}(\rho) || {\cal N}(\gamma)) - S(\rho || \gamma) \right],
 \end{aligned}
$$
by taking $G = \log \rho$. Combining these results altogether, we finally get the monotonicity of quantum relative entropy
$
S(\rho || \gamma) - S({\cal N}(\rho) || {\cal N}(\gamma)) \geq 0
$
as a consequence of the integral QFT.\hfill $\qed$

\section{Proof of Theorem~\ref{CovRevThm}}
The reversibility condition for the QFT can be proven by using the following relationship between the relative entropy and reversibility of a quantum channel \cite{Junge15}:
\begin{equation}
\label{RelEntRevInEq}
\langle \sigma \rangle \geq - \int_{-\infty}^\infty d\theta g_0(\theta) \log \left[ {\cal F}(\rho, ({\cal R}_\gamma^{\theta/2} \circ {\cal N})(\rho)) \right],
\end{equation}
where $\langle \sigma \rangle = S(\rho || \gamma) - S({\cal N} (\rho) || {\cal N} (\gamma))$. We also note that the Sutter-Berta-Tomamichel inequality in Eq.~(\ref{BertaIneq}) also leads to an alternative proof of the inequality given by Eq.~(\ref{RelEntRevInEq}). In order to see this, we take $p=1$ and $H'_1 = \frac{1}{2} U^\dagger [ \log {\cal N}(\rho) \otimes {\mathbb 1}_E] U$, $H'_2 = -\frac{1}{2} U^\dagger [ \log {\cal N}(\gamma) \otimes {\mathbb 1}_E] U$, $H'_3 = \frac{1}{2} \log \gamma$, $H'_4 = -\frac{1}{2} \log \rho$, and $H'_5 = \log \rho$. By using a similar logic to the proof of Theorem~\ref{2ndLawThm}, we have ${\rm (RHS)} = (1/2) \int_{-\infty}^\infty dt g_0(\theta) \log \left[ {\cal F}(\rho, ({\cal R}_\gamma^{\theta/2} \circ {\cal N})(\rho)) \right]$ and ${\rm (LHS)} \geq (1/2)\left[ S({\cal N}(\rho) || {\cal N}(\gamma)) - S(\rho || \gamma) \right] = -\langle \sigma \rangle /2$, which completes the proof.

Now we prove the Theorem. When $\langle \sigma \rangle = 0$, every ${\cal F}(\rho, ({\cal R}^{\theta/2}_\gamma \circ {\cal N})(\rho))$ should be $1$, which is the perfect reversibility condition. Conversely, if one of $({\cal R}^{\theta/2}_\gamma \circ {\cal N})(\rho) = \rho$, $\langle \sigma \rangle = S({\cal N}(\rho) \| {\cal N} (\gamma)) - S(\rho \| \gamma) =  S({\cal N}(\rho) \| {\cal N} (\gamma)) -  S( ({\cal R}_\gamma^{\theta/2} \circ {\cal N}) (\rho) \| ({\cal R}_\gamma^{\theta/2} \circ {\cal N}) (\gamma)) \leq 0$ by the monotonicity of quantum relative entropy. This condition implies that $\langle \sigma \rangle = 0$. Therefore, $\langle \sigma \rangle = 0$ \textit{if and only if} ${\cal F}(\rho, ({\cal R}^{\theta/2}_\gamma \circ {\cal N})(\rho)) = \rho$ for every $\theta$.

The bound for the recovery fidelity,
$$
\langle \sigma \rangle \geq - \log {\cal F}(\rho, (\bar{\cal R}_\gamma \circ {\cal N})(\rho)),
$$
is then obtained by the joint concavity of the fidelity function and the concavity of the logarithm function.  \hfill \qed


The Corollaries can be proved as follows:
For thermodynamic processes ${\cal N}_{\rm th}$, we note that ${\cal N}_{\rm th}(\gamma) = \gamma'$, where $\gamma = e^{-\beta H_S}/Z_S$ and $\gamma' = e^{-\beta H'_S}/Z'_S$.
Then the explicit form of this Petz recovery map is given by ${\cal R}_\gamma(\rho') = {\rm Tr}_B [ U^\dagger (\rho' \otimes \gamma'_B) U]$, which is a reverse thermodynamic process satisfying $U^\dagger ( H'_S +  H'_B) U = H_S + H_B$. We note that the thermodynamic process ${\cal R}_\gamma$ is covariant with respect to the reference state $\gamma$; thus, every rotated Petz recovery map ${\cal R}_\gamma^\theta$ reduces into the single form ${\cal R}_\gamma$. Then we have $\bar{\cal R}_\gamma = {\cal R}_\gamma$ and $\langle \sigma \rangle = \beta( \Delta F_{\rm eq} - \Delta F)$, and Theorem \ref{CovRevThm} leads to the reversibility condition.

For covariant processes, the only thing we need to show is that ${\cal R}_{{\cal D}(\rho)}$ is a covariant quantum channel and apply Theorem~\ref{CovRevThm}. This condition can be proven as both ${\cal D}(\rho)$ and ${\cal D} ({\cal N}_{\rm cov}(\rho))$ commute with the generator $L$. The rotated Petz recovery map is written as ${\cal R}_{{\cal D}(\rho)}^\theta = \JJ{{\cal D}(\rho)}{\frac{1}{2} + i \theta} \circ {\cal N}_{\rm cov}^\dagger \circ \JJ{{\cal D}({\cal N}_{\rm cov}(\rho))}{-\frac{1}{2} - i \theta}$, then the covariance condition is achieved by
$$
\begin{aligned}
{\cal R}_{{\cal D}(\rho)}^\theta& ( e^{- i L t} \chi e^{i L t}) \\
&= (\JJ{{\cal D}(\rho)}{\frac{1}{2} + i \theta} \circ {\cal N}_{\rm cov}^\dagger \circ \JJ{{\cal D}({\cal N}_{\rm cov}(\rho))}{-\frac{1}{2} - i \theta} \circ \JJ{\exp[L]}{-it}) \chi\\
&=  (\JJ{\exp[L]}{-it} \circ \JJ{{\cal D}(\rho)}{\frac{1}{2} + i \theta} \circ {\cal N}_{\rm cov}^\dagger \circ \JJ{{\cal D}({\cal N}_{\rm cov}(\rho))}{-\frac{1}{2} - i \theta}) (\chi) \\
&=  e^{-i L t} ( {\cal R}_{{\cal D}(\rho)}^\theta ( \chi )) e^{ i L t}
\end{aligned}
$$
as all the operations $\JJ{{\cal D}(\rho)}{\frac{1}{2} + i \theta} $, ${\cal N}_{\rm cov}^\dagger$, and $\JJ{{\cal D}({\cal N}_{\rm cov}(\rho))}{-\frac{1}{2} - i \theta} $ commute with $\JJ{\exp[L]}{-it}$. Therefore $\bar{\cal R}_{{\cal D}(\rho)}$ is covariant with respect to $L$ as it is a convex sum of the covariant operations ${\cal R}^\theta_{{\cal D}(\rho)}$. We also note that the recovery map $\bar{\cal R}_{{\cal D}(\rho)}$ fully recovers the diagonal elements:
$$
\begin{aligned}
({\cal D} \circ \bar{\cal R}_{{\cal D}(\rho)}) ({\cal N}_{\rm cov} (\rho)) &= ( \bar{\cal R}_{{\cal D}(\rho)} \circ {\cal D}) ({\cal N}_{\rm cov} (\rho))\\
&= \bar{\cal R}_{{\cal D}(\rho)} ( {\cal D} ({\cal N}_{\rm cov} (\rho)))\\
&= {\cal D}(\rho)
\end{aligned}
$$
as $\bar{\cal R}_{{\cal D}(\rho)}$ fully recovers ${\cal D} ({\cal N}_{\rm cov} (\rho))$ into ${\cal D}(\rho)$.

In a similar way to the covariant process, the recovery condition for the LOCC protocol can be proven by noting that the reverse process is another LOCC protocol as the reference state $\mathbb{1}_A \otimes \rho_B$ and its evolution ${\cal N}_{\rm LOCC} ( {\mathbb 1}_A \otimes \rho_B) =  {\mathbb 1}_A \otimes \sum_m P_m \rho_B^m \otimes \ket{m}_M \bra{m}$ act only on the side of $B$. Furthermore, we note that the projection onto the initial state is given by
$\ket{\Psi}_{AB} \bra{\Psi}$, so that $\kappa_\theta = \Tr [ \ket{\Psi}_{AB} \bra{\Psi}{\cal R}_{\mathbb{1}_A \otimes \rho_B}^{\theta/2} \circ {\cal N}_{\rm LOCC}) (\ket{\Psi}_{AB} \bra{\Psi}) ] = {\cal F}_\theta$, which is the recovery fidelity.
\hfill $\qed$

\section{The proof of Theorem~\ref{CohMerg}:}
By using the property of the covariant quantum channel ${\cal N}_{\rm cov}$ with respect to the generator $L = \sum_i \lambda_i \ket{i} \bra{i}$, the transition matrix $T_{ij \rightarrow k'l'}$ has a non-vanishing value only if  $\lambda_i - \lambda_j = \lambda_{k'} - \lambda_{l'}$. If the reference state $\gamma$ is taken to commute with the generator $L$, $\gamma$ and ${\cal N}_{\rm cov}(\gamma)$ have the same set of the eigenstates $\{ \ket{i} \}$ as the generator $L$. When an initial quantum state $\rho = \sum_{i,j} \rho_{ij} \ket{i} \bra{j}$ evolves into ${\cal N}_{\rm cov}(\rho) = \sum_{k',l'} {\cal N}_{\rm cov}(\rho)_{k'l'} \ket{k'}\bra{l'}$, the value of the off-diagonal element can be written as
$$
\begin{aligned}
\left| {\cal N}_{\rm cov}(\rho)_{k'l'} \right| &= \Big| \sum_{i,j} \rho_{ij} T_{ij \rightarrow k'l'} \Big|\\
&\leq \Big| \sum_{\Omega_{k'l'}^+} \rho_{ij} T_{ij \rightarrow k'l'} \Big|+ \Big| \sum_{\Omega_{k'l'}^-} \rho_{ij} T_{ij \rightarrow k'l'} \Big|,
\end{aligned}
$$
by using the triangle inequality.
Here, $\Omega_{k'l'} = \{ (i,j) | \lambda_i - \lambda_j = \lambda_{k'} - \lambda_{l'} \}$ denotes the set of $(i,j)$ having the non-vanishing transition matrix $T_{ij \rightarrow k'l'}$ for a given $(k', l')$, and $\Omega_{k'l'}^+$ and $\Omega_{k'l'}^-$ are the subsets of $\Omega_{k'l'}$ with the real part of information exchange being $(\delta q_R)_{ij \rightarrow k'l'} \geq 0 $ and $(\delta q_R)_{ij \rightarrow k'l'} < 0 $. From the fluctuation relation
${T_{ij \rightarrow k'l'}} {} = e^{ - \delta q_{ij \rightarrow k'l'} + i  \theta (\omega_{ij} - \omega'_{k'l'})} (\tilde T^{\theta}_{ij \leftarrow k'l'})^*$, we then obtain
$$
\begin{aligned}
&\left| {\cal N}_{\rm cov}(\rho)_{k'l'} \right|\\
&\quad \leq \Big| \sum_{\Omega_{k'l'}^+} \rho_{ij}  e^{ - \delta q_{ij \rightarrow k'l'} + i  \theta (\omega_{ij} - \omega'_{k'l'})} (\tilde{T}^{\theta}_{ij \leftarrow k'l'})^*    \Big| \\
&\qquad\qquad + \Big| \sum_{\Omega_{k'l'}^-} \rho_{ij} T_{ij \rightarrow k'l'} \Big|\\
&\quad \leq \sum_{\Omega_{k'l'}^+} | \rho_{ij} | e^{-(\delta q_R)_{ij \rightarrow k'l'}}  \Big| \tilde{T}^\theta_{ij \leftarrow k'l'}  \Big| +  \sum_{\Omega_{k'l'}^-} | \rho_{ij} | \Big| T_{ij \rightarrow k'l'} \Big|\\
&\quad \leq \sum_{\Omega_{k'l'}^+} | \rho_{ij} | e^{-(\delta q_R)_{ij \rightarrow k'l'}}  +  \sum_{\Omega_{k'l'}^-} | \rho_{ij} |,
\end{aligned}
$$
by noting that both $\big| T_{ij  \rightarrow k'l'} \big|$ and $\big| \tilde{T}^\theta_{ij \leftarrow k'l'}   \big|$ are less than $1$.\hfill $\qed\\$

\end{document}